\documentclass[12pt]{article}
\usepackage{amsfonts,amssymb,amsmath}
\usepackage{color}
\usepackage{pst-grad, pst-node}
\usepackage{rotating}
\usepackage{epsfig}
\usepackage{graphicx}
\usepackage{graphics}

\def\Box{\leavevmode\vbox{\hrule
     \hbox{\vrule\kern4pt\vbox{\kern4pt}%
           \vrule}\hrule}}
\def\blackbox{\leavevmode\vrule height 5pt width 4pt depth 0pt\relax}
\def\endproof{\null\hfill {$\blackbox$}\bigskip}

\newcounter{appendix}
\def\appendix{\advance\c@appendix by 1
   \def\thesection{\Alph{section}}
   \ifnum\c@appendix=1 \setcounter{section}{-1} \fi
   \@startsection {section}{1}{\z@}{-3.5ex plus -1ex minus 
   -.2ex}{2.3ex plus .2ex}{\Large\bf}}

\def\paragraph#1{{\bf #1\ }}

\newtheorem{lemma}{Lemma}[section]  

\newtheorem{theorem}[lemma]{Theorem}

\newtheorem{definition}[lemma]{Definition}

\newtheorem{proposition}[lemma]{Proposition}

\title{Diffusion in a continuum model of self-propelled particles with alignment interaction} 
\author{P. Degond $^{(1,2)}$, T. Yang$^{(3)}$} 
\date{} 
\begin{document}

\maketitle

\vspace{0.5 cm}

\begin{center}
1-Université de Toulouse; UPS, INSA, UT1, UTM ;\\ 
Institut de Mathématiques de Toulouse ; \\
F-31062 Toulouse, France. \\
2-CNRS; Institut de Mathématiques de Toulouse UMR 5219 ;\\ 
F-31062 Toulouse, France.\\
email: pierre.degond@math.univ-toulouse.fr \\
\end{center}

\begin{center}
3- Department of Mathematics, \\
City University of Hong Kong, \\ 
83 Tat Chee Avenue, Kowloon, Hong Kong 
\\
email: matyang@cityu.edu.hk \\
\end{center}

\begin{abstract}
In this paper, we provide the $O(\varepsilon)$ corrections to the hydrodynamic model derived by Degond \& Motsch from a kinetic version of the model by Vicsek \& coauthors describing flocking biological agents. The parameter $\varepsilon$ stands for the ratio of the microscopic to the macroscopic scales. The $O(\varepsilon)$ corrected model involves diffusion terms in both the mass and velocity equations as well as terms which are quadratic functions of the first order derivatives of the density and velocity. The derivation method is based on the standard Chapman-Enskog theory, but is significantly more complex than usual due to both the non-isotropy of the fluid and the lack of momentum conservation. 
\end{abstract}

\medskip
\noindent
{\bf Acknowledgements:} This work was supported by the Marie Curie Actions of the European 
Commission in the frame of the DEASE project (MEST-CT-2005-021122) and by the french 'Agence Nationale pour la Recherche (ANR)' in the frame of the contract 'Panurge' (ANR-07-BLAN-0208-03).


\vskip 0.4cm


\setcounter{equation}{0}
\section{Introduction}
\label{intro}

This paper is a development of a previous work \cite{Degond_CRAS07,Degond_M3AS08} about continuum models of self-propelled particles subject to alignment interaction. This class of models describes swarming behaviour among biological species and attempts at providing a simplified theoretical framework to experimental observations (see recent observations in Refs. \cite{Ballerini_PNAS08,Becco_PhysicaA06,Buhl_Science06,Gautrais_JMB09,Ingham_BMCmicrobio08}). 

The starting point is a particle model (or Individual-Based Model (IBM)) discussed in e.g. Refs. \cite{Aoki_BJSSF82,Couzin_JTheoBio02,Hemelrijk_BehavEcology04,Mirabet_EcoMod07}, where the interaction between biological agents such as fish or birds is described by three interaction ranges: a close range where repulsion occurs, a long range where attraction prevails and a medium range where agents tend to align with each other. In the Vicsek model,\cite{Vicsek_PRL95} the alignment interaction is singled out and analyzed. More precisely, each particle moves with a constant and uniform speed and aligns with the average direction of all neighbours within an interaction distance $R$, up to some angular fluctuation. Vicsek and co-authors\cite{Vicsek_PRL95} show that phase transitions from disorder to order appear as the noise intensity decreases or the density increases. This model has triggered a wealth of publications\cite{Aldana_JSP03,Barbaro_MCS09,Chate_PhysRevE08,Czirok_PhysicaA99,Gautrais_AZF08,Gregoire_PRL04} and given rise to various variants\cite{Chate_PRL06,Szabo_PhysRevE09}. 

The task of deriving kinetic (Boltzmann-like) or continuum (fluid-like) models from this model has been undertaken by various approaches (see Refs. \cite{Bertin_PhysRevE06,Peruani_EPJST08} for kinetic models and Refs. \cite{Bertin_PhysRevE06,Czirok_PRL99,Czirok_PhysicaA00,Peruani_EPJST08,Ratushnaya_PhysicaA07} for fluid models). However, these models are based on physical arguments and the mathematical approach of Ref. \cite{Degond_CRAS07,Degond_M3AS08} leads to a different type of model: indeed, this model is not of diffusive nature like in Refs. \cite{Bertin_PhysRevE06,Czirok_PRL99,Czirok_PhysicaA00,Peruani_EPJST08} and is local, by contrast to Ref. \cite{Ratushnaya_PhysicaA07}. The differences arise because the model of Refs. \cite{Degond_CRAS07,Degond_M3AS08} is derived in the large time and space scale limit in which, at leading order, the interactions are local and the diffusivities, negligible.  

The goal of this paper is to investigate how the model of Ref. \cite{Degond_CRAS07,Degond_M3AS08} must be adapted to account for small but finite nonlocality and diffusivity. We will see that the resulting model involves complex effects due to the non-isotropy of the fluid (by contrast to standard fluids). This point will be further developed in section \ref{sec_position}. Similar to the rarefied gas dynamics case (see e.g. Refs. \cite{Caflisch_CPAM80,CIP,degond03:_macros_boltz,Sone}), the derivation is based on the Chapman-Enskog expansion method. However, in the present case, the computations are significantly more complex because of the non-isotropy of the fluid and of the lack of momentum conservation. 

The model derived in Ref. \cite{Degond_CRAS07,Degond_M3AS08}  has been extended in Ref. \cite{Frouvelle_preprint} to account for anisotropic vision and density-dependent interaction frequency. Its numerical resolution is performed in Ref. \cite{Motsch_preprint}.

Other Individual-Based Models involving attraction-repulsion interactions can be found in Refs.  \cite{Dorsogna_PRL06,Forgoston_PhysRevE08,Mogilner_JMB03,Peruani_PhysRevE06} and continuum models, in Refs.  \cite{Bertozzi_Nonlinearity09,Carrillo_KRM09,Chuang_PhysicaD07,Csahok_PhysicaA97,Degond_JSP08,Mogilner_JMathBio99,Topaz_SIAP04,Topaz_BullMathBio06}. The existence of flocking or non-flocking behaviour for the Cucker-Smale model\cite{Cucker_IEEEAutomCont07,Cucker_JMPA08,Cucker-M3AS09} (which is similar to the Vicsek model but without noise nor speed constraint) has received a great deal of attention\cite{Carrillo_SIMA10,Duan_preprint,Ha_KRM08,Ha_CMS09,Ha_CMS09_2,Ha_JDynDiff09,Shen_SIAP07}.


\setcounter{equation}{0}
\section{Position of the problem and main result}
\label{sec_position}

In Refs. \cite{Degond_CRAS07,Degond_M3AS08} the following continuum model of self-driven particles with alignment interaction has been derived: 
\begin{eqnarray} 
& & \partial_t \rho + \nabla \cdot (c_1 \rho \Omega)  = 0,
\label{mass_eq_0} \\
& &  \rho \, \left( \partial_t \Omega + c_2 (\Omega \cdot \nabla) \Omega \right) +  c_3  \,  (\mbox{Id} - \Omega \otimes \Omega) \nabla \rho = 0,
\label{Omega_eq_0} 
\end{eqnarray}
where $\rho = \rho (x,t)$ is the number density of the particles and $\Omega(x,t)$ is the direction of their average velocity, which satisfies $|\Omega|=1$. $c_1$, $c_2$ and $c_3$ are constants which are computed from the underlying microscopic dynamics, where $c_2 < c_1 < 1$ and $c_1$, $c_2$, $c_3 >0$. The matrix $(\mbox{Id} - \Omega \otimes \Omega)$ denotes the orthogonal projection matrix onto the plane orthogonal to $\Omega$. The notations Id and $\otimes$ respectively refer to the identity matrix and to the tensor product. 

The derivation of this model, according to Ref. \cite{Degond_M3AS08}, proceeds through several (formal) asymptotic limits. The starting point is a time-discrete particle model proposed by Vicsek and co-authors\cite{Vicsek_PRL95}. In the Vicsek model,  the particles move with a constant speed and, at discrete times, align their velocities to the mean velocity of their neighbours, up to some small noise. In Ref. \cite{Degond_M3AS08}, a continuous in time version of this particle model is first proposed in which the alignment interaction is modeled through a relaxation term and the noise, by a Brownian motion on the particle velocities. The formal mean-field limit of this time-continuous particle model (as the number of particles tends to infinity) leads to the following nonlinear Fokker-Planck model:
\begin{eqnarray} 
& & \hspace{-1cm} \partial_t f + \omega \cdot \nabla f + \nabla_\omega \cdot (F f) = d \Delta_\omega f ,  
\label{FP_mf}  \\
& & \hspace{-1cm} F (x, \omega, t) = \nu(\cos \bar \theta) \, (\mbox{Id} - \omega \otimes \omega) 
\bar \omega (x,\omega,t) , 
\label{Force_mf} \\
& & \hspace{-1cm} \bar \omega (x, \omega, t) = \frac{J(x,t)}{|J(x,t)|}, \quad J(x,t) =  \int_{ y \in {\mathbb R}^3 , \, \upsilon \in {\mathbb S}^2 } K(|x-y|) \, 
\upsilon \, f(y, \upsilon,t) \, dy \, d\upsilon \, \, . 
\label{bar_omega_mf}
\end{eqnarray}
Here $f(x,\omega, t)$ is the particle distribution function depending on the space variable $x \in {\mathbb R}^3$, the velocity direction $\omega \in {\mathbb S}^2$ and the time $t$. $d$ is a scaled diffusion constant associated with the Brownian noise, and $F (x, \omega, t)$ is the mean-field interaction force between the particles which depends on an interaction frequency $\nu$. This force tends to align the particles to the direction $\bar \omega$ which is the direction of the particle flux $J$ in a neighbourhood of $x$ weighted by the kernel $K$. Typically, if $K$ is the indicator function of the ball centered at $0$ and of radius $R$ then $J$ is the particle flux integrated over a ball centered at $x$ and of radius $R$. The matrix $(\mbox{Id} - \omega \otimes \omega) $ is the projection matrix onto the normal plane to $\omega$ and we assume that the collision frequency may depend on $\cos \bar \theta = (\omega \cdot \bar \omega)$, i.e. on the cosine of the angle between $\omega$ and $\bar \omega$. 

\medskip
\noindent
{\bf Notation convention:} The $\nabla$ (and below, $\Delta$) symbols indicate the nabla and Laplacian operators with respect to $x$ while $\nabla_\omega$ and $\Delta_\omega$ denote the nabla and Laplace-Beltrami operators with respect to $\omega$. Expressions of $\nabla_\omega$ and $\Delta_\omega$ in spherical coordinates will be recalled later. 

\medskip
System (\ref{FP_mf})-(\ref{bar_omega_mf}) is written in a scaled form: the time and space scales have been chosen such that the particle speed $|\omega|$ is exactly $1$ and that both $d$ and $\nu$ are of order unity. With these so-called microscopic scales, the typical time and distance between two particle interactions are both $O(1)$. We refer to Ref. \cite{Degond_M3AS08} for a discussion of this point. 

By contrast, model (\ref{mass_eq_0}), (\ref{Omega_eq_0}) is designed to capture the large scale dynamics only, while averaging out the microscopic scales. Therefore, the passage from model (\ref{FP_mf})-(\ref{bar_omega_mf}) to model (\ref{mass_eq_0}), (\ref{Omega_eq_0}) requires a change of scales. Let $\varepsilon \ll 1$ be a measure of the ratio of the microscopic length scale to the size of the  observation domain. Here, the relevant scaling is a hydrodynamic scaling, which means that $\varepsilon$ is also equal to the ratio of the microscopic time scale to the macroscopic observation time. To express model (\ref{FP_mf})-(\ref{bar_omega_mf}) in terms of the macroscopic time and  space scales, we perform the change of variables $\tilde x = \varepsilon x$, $\tilde t = \varepsilon t$. In doing so, we must assign a scale to the interaction kernel $K$ (i.e. to the interaction range $R$). A key assumption in the present work, following Ref. \cite{Degond_M3AS08}, is that this interaction range is microscopic, and therefore of order $\varepsilon$ when using the macroscopic scales. This change of variables leads to (dropping the tildes for clarity):
\begin{eqnarray} 
& & \hspace{-1cm} \varepsilon ( \partial_t f^\varepsilon + \omega \cdot \nabla f^\varepsilon ) = -  \nabla_\omega \cdot (F^\varepsilon f^\varepsilon) + d \Delta_\omega f^\varepsilon ,  
\label{FP_mf_large} \\
& & \hspace{-1cm} F^\varepsilon (x, \omega, t) = \nu (\cos \theta^\varepsilon) \, \, (\mbox{Id} - \omega \otimes \omega) 
\bar \omega^\varepsilon , \quad \cos \theta^\varepsilon = \omega \cdot \bar \omega^\varepsilon, 
\label{Force_mf_large} \\
& & \hspace{-1cm} \bar \omega^\varepsilon (x,  t) = \frac{J^\varepsilon(x,t)}{|J^\varepsilon(x,t)|}, \quad J^\varepsilon(x,t) = \int_{ y \in {\mathbb R}^3 , \, \upsilon \in {\mathbb S}^2 } K \left( \left| \frac{x-y}{\varepsilon} \right| \right) \, 
\upsilon \, f^\varepsilon(y, \upsilon,t) \, dy \, d\upsilon  \, . 
\label{bar_omega_mf_large} 
\end{eqnarray}
It is an easy matter to see that $\bar \omega^\varepsilon$ has the following expansion:
\begin{eqnarray} 
& & \hspace{-1cm} \bar \omega^\varepsilon = \Omega^\varepsilon + \varepsilon^2 \bar \omega^\varepsilon_2 
+ O(\varepsilon^4)
\, ,
\label{bar_omega_mf_expan} \\
& & \hspace{-1cm} \Omega^\varepsilon (x, t) = \frac{j^\varepsilon(x,t)}{|j^\varepsilon(x,t)|}, \quad j^\varepsilon(x,t) = \int_{\upsilon \in {\mathbb S}^2 } \upsilon \, f^\varepsilon(x, \upsilon,t) \, d\upsilon  \, , 
\label{Omega} \\
& & \hspace{-1cm} \bar \omega^\varepsilon_2 = {\mathcal K} (\mbox{Id} - \Omega^\varepsilon \otimes \Omega^\varepsilon) \frac{\Delta j^\varepsilon}{|j^\varepsilon|}  \, . 
\label{bar_omega_2} 
\end{eqnarray}
$\Omega^\varepsilon$ is the direction of the local flux $j^\varepsilon$ and the constant ${\mathcal K}$ depends on the interaction kernel $K$ through: 
$$ {\mathcal K} = \frac{K_2}{6 K_0}, \quad K_p = \int_{{\mathbb R}^3} K(|\xi|) \, |\xi|^p \, d \xi \, . $$
Accordingly, $F^\varepsilon$ can be expanded: 
\begin{eqnarray} 
& & \hspace{-1cm} F^\varepsilon = F_0^\varepsilon + \varepsilon^2 F_2^\varepsilon + O(\varepsilon^4)  \, , \quad F_2^\varepsilon = F_2^{\varepsilon 1} + F_2^{\varepsilon 2} , \label{Fepsilon_expan} \\
& & \hspace{-1cm} F_0^\varepsilon = \nu (\cos \theta^\varepsilon) (\mbox{Id} - \omega \otimes \omega) \Omega^\varepsilon  , \quad \cos \theta^\varepsilon = \omega \cdot \Omega^\varepsilon, 
\label{F0_expan} \\
& & \hspace{-1cm} 
F_2^{\varepsilon 1} = \nu(\cos \theta^\varepsilon) (\mbox{Id} - \omega \otimes \omega) \bar \omega^\varepsilon_2, \quad 
F_2^{\varepsilon 2} = \nu'(\cos \theta^\varepsilon) \, (\omega \cdot \bar \omega^\varepsilon_2) \,  (\mbox{Id} - \omega \otimes \omega) \Omega^\varepsilon
 \, , 
\label{F2_expan}  
\end{eqnarray}
where $\nu'(\cos \theta)$ is the derivative of $\nu(\cos \theta)$ with respect to $\cos \theta$. Because of the dependence of $K$ upon the distance $|x-y|$ only, all odd powers of $\varepsilon$ vanish in the expansion. This would not be the case if we considered more general kernels such as those\cite{Frouvelle_preprint} depending on the angle between $\omega$ and $x-y$ . The consideration of more general kernels is left to future work. 

Consequently, we will consider the following expanded Fokker-Planck model
\begin{eqnarray} 
& & \hspace{-1cm}  \partial_t f^\varepsilon + \omega \cdot \nabla f^\varepsilon + \varepsilon \nabla_\omega \cdot (F_2^\varepsilon f^\varepsilon) = \frac{1}{\varepsilon} (-  \nabla_\omega \cdot (F_0^\varepsilon f^\varepsilon) + d \Delta_\omega f^\varepsilon ) + O(\varepsilon^2),  
\label{FP_mf_eps_0}
\end{eqnarray}
where the terms $F_0^\varepsilon$ and $F_2^\varepsilon$ are defined by (\ref{F0_expan}) and (\ref{F2_expan})
We note that, at leading order, the interaction force $F_0^\varepsilon$ only depends on the local flux $j^\varepsilon$ and that the corrections due to the nonlocality of the interaction force appear in the $O(\varepsilon)$ terms only. This is due to the assumption that the radius of the interaction region is very small (of order $\varepsilon$) in the macroscopic variables. 

In Ref. \cite{Degond_M3AS08}, it has been proved that model (\ref{mass_eq_0}), (\ref{Omega_eq_0}) is the formal hydrodynamic limit $\varepsilon \to 0$ of the mean-field model (\ref{FP_mf_eps_0}). Additionally, Ref. \cite{Degond_M3AS08} provides the connection between the coefficients $c_1$, $c_2$ and $c_3$ of the macroscopic model to the coefficients $\nu$ and $d$ of the microscopic one. The goal of this paper is to investigate what diffusive corrections are obtained when keeping the $O(\varepsilon)$ corrections in the Chapman-Enskog expansion of $f^\varepsilon$. These terms describe the response of the system to the appearence of gradients of the state variables $\rho$ and $\Omega$. Here, because the fluid is anisotropic, and has only invariance through rotations about $\Omega$, these gradients must be split into their components parallel and perpendicular to $\Omega$. 

To this aim, we denote by 
\begin{eqnarray}
{\mathcal O}_\bot = \mbox{Id} - \Omega \otimes \Omega, \quad {\mathcal O}_\parallel = \Omega \otimes \Omega,
\label{def_O}
\end{eqnarray}
the orthogonal projection matrices onto the plane normal to $\Omega$ and onto the line spanned by $\Omega$ respectively. For a given vector $X \in {\mathbb R}^3$, we recall that 
$$ {\mathcal O}_\bot X = X - (X \cdot \Omega) \Omega = \Omega \times (X \times \Omega), \quad {\mathcal O}_\parallel X = (X \cdot \Omega) \Omega. $$
Using these projections, any vector field $A$ and tensor field $B$ can be decomposed into parallel and transverse components according to:
\begin{eqnarray} 
& & \hspace{-1cm}   
A =  A_\bot + A_\parallel , \quad 
B = B_{\bot, \bot} + B_{\bot, \parallel} + B_{\parallel, \bot} + B_{\parallel, \parallel},
\label{def_decomp}
\end{eqnarray}
defined by
\begin{eqnarray*} 
& & \hspace{-1cm}   
A_\bot = {\mathcal O}_\bot A, \quad  A_\parallel = {\mathcal O}_\parallel  A , \\
& & \hspace{-1cm}   
B_{\bot, \bot} = {\mathcal O}_\bot B {\mathcal O}_\bot ,  \quad  B_{\bot, \parallel} = {\mathcal O}_\bot B {\mathcal O}_\parallel, \quad
B_{\parallel, \bot} = {\mathcal O}_\parallel B {\mathcal O}_\bot, \quad   B_{\parallel, \parallel}  = {\mathcal O}_\parallel B {\mathcal O}_\parallel.
\end{eqnarray*}

Now, we decompose gradient fields according to their parallel and normal components to $\Omega$. For a scalar function $f$, we define the normal and parallel gradients as
$$ \nabla_\bot f = (\nabla f)_\bot, \quad \nabla_\parallel f = (\nabla f)_\parallel. $$
Similarly, we may decompose the gradient of a vector field $u$ into 
$$ \nabla u =  \nabla_{\bot, \bot} u + \nabla_{\bot, \parallel} u + \nabla_{\parallel, \bot} u + \nabla_{\parallel, \parallel} u, $$ 
using the tensor decomposition (\ref{def_decomp}). Applying this decomposition to $\nabla \Omega$ itself, we find: 
\begin{eqnarray}
& & \hspace{-1cm}
\nabla_{\bot, \bot} \Omega = \nabla \Omega - \Omega \otimes (\Omega \cdot \nabla) \Omega\, , \quad \nabla_{\parallel, \bot} \Omega = \Omega \otimes (\Omega \cdot \nabla) \Omega, \label{decomp_nabla_Omega_1} \\
& & \hspace{-1cm}
\nabla_{\bot, \parallel} \Omega = 0 \, , \quad  \nabla_{\parallel, \parallel} \Omega = 0.
\label{decomp_nabla_Omega_2}
\end{eqnarray}
The last line is a consequence of $(\nabla \Omega) \Omega = 0$, which is found by taking the derivative of the relation $|\Omega|^2 = 1$. 

\begin{figure}
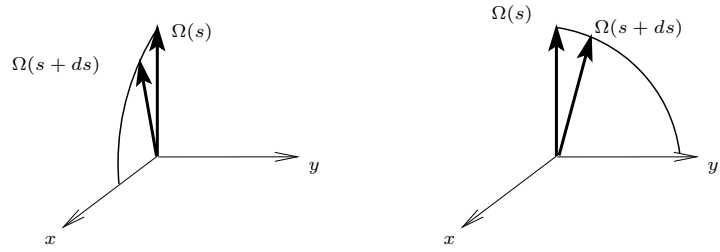

\begin{center}
\input{OmnaOmx.pstex_t} \hspace{1.5cm} 
\input{OmnaOmy.pstex_t}
\caption{$(\Omega \cdot \nabla) \Omega$ describes the rate of tilt of $\Omega$ as one moves from one point $s$ to the neighbouring point $s+ds$ on the flow lines. In the left figure, $(\Omega \cdot \nabla) \Omega$ will be aligned with the $x$ axis ; in the right figure, with the $y$ axis.}
\label{Fig_1}
\end{center}
\end{figure}

\begin{figure}
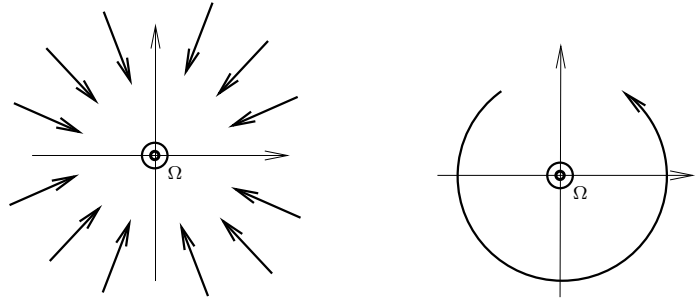

\begin{center}
\input{divOm.pstex_t} \hspace{1.5cm}
\input{Gamma.pstex_t}
\caption{$\nabla \cdot \Omega$ refers to converging (or diverging) flow patterns in the plane normal to $\Omega$ (left picture) while $\Gamma(\Omega)$ refers to swirling patterns around $\Omega$ (right picture). $\Omega$ is directed across the plane of the picture and pointing towards the observer (represented by the circled point at the origin). }	
\label{Fig_2}
\end{center}
\end{figure}

\begin{figure}
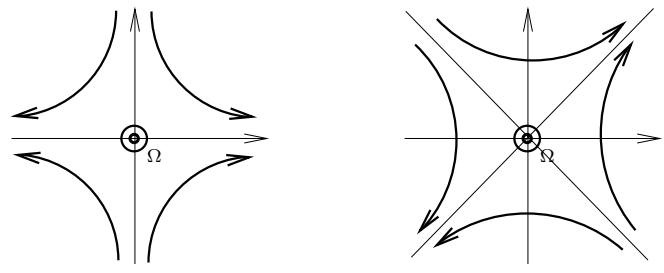

\begin{center}
\input{sigma1.pstex_t} \hspace{1.5cm} 
\input{sigma2.pstex_t}
\caption{$\sigma (\Omega)$ describes shear in the plane normal to $\Omega$. 
The left picture depicts the case associated to $\sigma_1$ and the right picture, the case associated to $\sigma_2$. $\Omega$ is directed across the plane of the picture and pointing towards the observer (represented by the circled point at the origin).}	
\label{Fig_3}
\end{center}
\end{figure}

In compressible Navier-Stokes equations\cite{degond03:_macros_boltz}, the diffusion terms can be expressed as functions of only two quantities constructed with the gradient of the velocity field $u$: the traceless rate of strain tensor $\sigma (u) = \nabla u + (\nabla u)^T - (2/3) (\nabla \cdot u) \mbox{Id}$, and the divergence field $\nabla \cdot u$ (the exponent $T$ denotes the matrix transpose). Here, the anisotropy of the problem gives rise to different diffusivities in the directions parallel or normal to $\Omega$ and we need to split the matrix $\nabla \Omega$ into a larger number of separate entities. To this aim, we note that 
\begin{eqnarray}
& & \hspace{-1cm}
\nabla \cdot \Omega = \mbox{Tr} (\nabla \Omega) = \mbox{Tr} (\nabla_{\bot, \bot} \Omega) , 
\label{div_Omega}
\end{eqnarray}
where 'Tr' denotes the trace of a tensor. The traceless tensor $\nabla_{\bot, \bot} \Omega - (1/2) (\nabla \cdot \Omega) {\mathcal O}_\bot$ is decomposed in its symmetric and anti-symmetric parts $\sigma(\Omega)$ and $\Gamma(\Omega)$: 
\begin{eqnarray}
\sigma(\Omega)& = &\nabla_{\bot, \bot} \Omega + (\nabla_{\bot, \bot} \Omega)^T - (\nabla \cdot \Omega) {\mathcal O}_\bot \nonumber \\
&= &{\mathcal O}_\bot \, ( \nabla \Omega + (\nabla \Omega)^T 
- (\nabla \cdot \Omega) \, \mbox{Id} ) \, {\mathcal O}_\bot,  \label{def_sigma} \\
\Gamma(\Omega) &= &\nabla_{\bot, \bot} \Omega - (\nabla_{\bot, \bot} \Omega)^T
= {\mathcal O}_\bot \, ( \nabla \Omega - (\nabla \Omega)^T ) \, {\mathcal O}_\bot .   \label{def_Delta} 
\end{eqnarray}
These relations will be used in the form: 
\begin{eqnarray}
\nabla_{\bot, \bot} \Omega &=& \frac{1}{2} (\sigma(\Omega)+ \Gamma(\Omega)) + \frac{1}{2} (\nabla \cdot \Omega) {\mathcal O}_\bot,  \label{nabla_botbot_Omega}\\
(\nabla_{\bot, \bot} \Omega)^T &=& \frac{1}{2} (\sigma(\Omega) - \Gamma(\Omega)) + \frac{1}{2} (\nabla \cdot \Omega) {\mathcal O}_\bot.  \label{nabla_botbot_Omega_transpose}
\end{eqnarray}
The non-zero block of $\sigma(\Omega)$ is a 2 by 2 symmetric traceless tensor and the non-zero block of $\Gamma(\Omega)$ is a 2 by 2 anti-symmetric tensor. We note that,  for a given vector $X \in {\mathbb R}^3$: 
\begin{eqnarray}
\Gamma (\Omega) X = - (\Gamma (\Omega))^T X = ((\nabla \times \Omega) \cdot \Omega) \, X \times \Omega.   
\label{Delta_Omega}
\end{eqnarray}
Similarly, through (\ref{decomp_nabla_Omega_1}), $\nabla_{\bot, \parallel} \Omega$ depends only on $(\Omega \cdot \nabla) \Omega$ and we have:
\begin{eqnarray}
(\Omega \cdot \nabla) \Omega = (\nabla \Omega)^T \Omega = (\nabla \times \Omega) \times \Omega,  
\label{Omega_grad_Omega}
\end{eqnarray}
where $\nabla \times \Omega$ denotes the curl of $\Omega$. Physically, $(\Omega \cdot \nabla) \Omega$ describes the rate of tilt of $\Omega$ as one moves along the flow lines (see figure \ref{Fig_1}). The other quantities describe elementary flow patterns in the plane normal to $\Omega$: $\nabla \cdot \Omega$  refers to convergent or divergent flows in the direction normal to $\Omega$ while $\Gamma(\Omega)$ refers to swirling patterns around $\Omega$ (see figure \ref{Fig_2}) and $\sigma(\Omega)$ to shear patterns with one converging and one diverging orthogonal directions. Restricted to the plane normal to $\Omega$, $\sigma(\Omega)$ is a symmetric traceless $2 \times 2$ matrix. Therefore, it can be expressed as a linear combination of the two elementary matrices: 
$$ \sigma_1 = \left( \begin{array}{cc} 1 & 0 \\ 0 & -1 \end{array} \right) , \quad 
\sigma_2 = \left( \begin{array}{cc} 0 & 1 \\ 1 & 0 \end{array} \right) ,$$
each corresponding to an elementary flow pattern (see figure \ref{Fig_3}). 

In the compressible Navier-Stokes equations, there are no diffusion terms in the mass  equation and diffusion terms in the energy equation are expressed in terms of the temperature gradient $\nabla T$ as a whole. Here, the temperature is constant (it is fixed by the noise level) but, because of the lack of momentum conservation, the diffusion terms in the mass conservation equation are not zero. Additionally, both the mass and velocity diffusions depend on $\nabla \rho$ as well as on $\nabla \Omega$. Therefore, similar to $\nabla \Omega$, we decompose $\nabla \rho$ into its parallel and normal components (respectively $(\Omega \cdot \nabla) \rho$ and $\nabla_\bot \rho$). 

Finally, the diffusion terms are composed of two parts: the first one is a quadratic form of the gradients in the set $\{ \nabla_\bot \rho, \, (\Omega \cdot \nabla) \rho, \, (\Omega \cdot \nabla) \Omega, \, \sigma(\Omega), \, \Gamma(\Omega), \, \nabla \cdot \Omega \}$ with coefficients depending on $(\rho,\Omega)$ ; the second one is a linear combination of second derivatives constructed by taking parallel derivatives $\Omega \cdot \nabla$ or perpendicular derivatives $\nabla_\bot$ of the gradients in the above list. This is precisely stated in the following theorem, which constitutes the main result of this paper:  

\begin{theorem} (formal)
The following model 
\begin{eqnarray} 
& & \partial_t \rho + \nabla \cdot (c_1 \rho \Omega)  =\varepsilon R_1, 
\label{mass_eq_eps} \\
& &  \rho \, \left( \partial_t \Omega + c_2 (\Omega \cdot \nabla) \Omega \right) +  c_3  \,   \nabla_\bot \rho = \varepsilon R_2, 
 \label{Omega_eq_eps} 
\end{eqnarray}
where $R_1$ and $R_2$ given below, provides a second order approximation of the moments of the solution $\rho^\varepsilon$ and $\Omega^\varepsilon$ of the initial model (\ref{FP_mf_eps_0}).  The right-hand-sides are given by: 
\begin{eqnarray}
& & \hspace{-1cm}  
R_1 = \beta \nabla \cdot ( (\Omega \cdot \nabla \rho) \Omega) + \gamma \nabla \cdot ( \rho (\nabla \cdot \Omega) \Omega) , \label{express_rhs_FP_rho} \\
& & \hspace{-1cm}  
R_2 = {\mathcal Q} + {\mathcal D} , \label{express_rhs_FP_Omega} \\
& & \hspace{-1cm}  
{\mathcal Q} = 
\, + \, {\mathcal Q}_1 \,   (\Omega \cdot \nabla \rho) \nabla_\bot \rho
\, + \, {\mathcal Q}_2 \, (\nabla \cdot \Omega) \nabla_\bot \rho 
\, + \, {\mathcal Q}_3 \, \sigma (\Omega) \nabla_\bot \rho 
\nonumber \\
& & \hspace{-0.5cm}
\, + \, {\mathcal Q}_4  \, \Gamma (\Omega) \nabla_\bot \rho 
\, + \, {\mathcal Q}_5  \, (\Omega \cdot \nabla \rho) (\Omega \cdot \nabla) \Omega
\, + \, {\mathcal Q}_6  \, (\nabla \cdot \Omega) (\Omega \cdot \nabla) \Omega
\nonumber \\
& & \hspace{-0.5cm}
\, + \, {\mathcal Q}_7  \, \sigma (\Omega) (\Omega \cdot \nabla) \Omega
\, + \, {\mathcal Q}_8  \, \Gamma (\Omega) (\Omega \cdot \nabla) \Omega
, \label{mathcal_Q} \\
& & \hspace{-1cm}  
{\mathcal D} = {\mathcal D}_1  \, {\mathcal O}_\bot (\Omega \cdot \nabla) \nabla_\bot \rho
\, + \, {\mathcal D}_2  \, {\mathcal O}_\bot (\Omega \cdot \nabla)( (\Omega \cdot \nabla) \Omega)   
\, + \, {\mathcal D}_3  \, \nabla_\bot (\nabla \cdot \Omega)   
\nonumber \\
& & \hspace{-0.5cm}
\, + \, {\mathcal D}_4  \, {\mathcal O}_\bot \nabla \cdot \sigma (\Omega)   
\, + \, {\mathcal D}_5  \, {\mathcal O}_\bot \nabla \cdot \Gamma (\Omega)   , \label{mathcal_D}
\end{eqnarray}
where $\beta$, $\gamma$, ${\mathcal Q}_j$ for $j=1, \ldots ,  8$,  ${\mathcal D}_j$ for $j=1, \ldots ,  5$ are coefficients, possibly depending on $\rho$, which are given below.  Additionally, we have $\beta >0$. 
\label{theo_limit}
\end{theorem}

The structure of $R_2$ is as announced: it is decomposed into a term ${\mathcal Q}$ which is a quadratic function of the gradients and a term ${\mathcal D}$ which consists of derivatives of these gradients. Both terms involve coefficients which may depend on $\rho$. The quadratic part combines products of the parallel gradient of $\rho$, $\Omega \cdot \nabla \rho$, with the perpendicular gradient of $\rho$, $\nabla_\bot \rho$, and similarly for $\Omega$ (the parallel gradient of $\Omega$ being $\Omega \cdot \nabla \Omega$, the perpendicular ones being defined as any of gradients in the list $\{ \nabla \cdot \Omega, \sigma(\Omega), \Gamma(\Omega) \}$) and products of the perpendicular gradient of $\rho$ with the  perpendicular gradients of $\Omega$ or parallel gradients of $\rho$ with parallel gradients of $\Omega$. The diffusive part involves only parallel gradients of the parallel gradient of $\Omega$, or perpendicular gradients of perpendicular gradients of $\Omega$, and finally a parallel gradient of a perpendicular gradient of $\rho$. In spite of its complex expression, $R_2$ has a lot of structure, since a general function of this form would have 21 different terms in the quadratic part and 10 in the diffusion part, instead of respectively 8 and 5. 

The main objective of this paper is the proof of this theorem. In future work, the properties of this system will be analyzed. In particular, the question of the well-posedness of the system under the sole condition $\beta >0$ will be investigated, at least on a simpler model. Indeed, the
other  diffusion coefficients have no definite signs, but the constraint $|\Omega| = 1$ should prevent the formation of singularities, by contrast to the usual backwards heat equation.  

The organization of the proof is as follows. In section \ref{sec_propQ}, we recall the main properties of the collision operator $Q$ that are proved in Ref. \cite{Degond_M3AS08} and provide additional properties of the linearized operator $Q$ about an equilibrium. This prepares the terrain for the Chapman-Enskog expansion which is performed in section \ref{sec_CE}. We start section \ref{sec_CE} by an exposition of the main steps to be accomplished. Then, we successively examine the solvability condition for the existence of the first order correction to the equilibrium, the finding of an analytical expression of it as a function of elementary solutions of the linearized collision operator, and finally, the computations of the moments of this correction which precisely give rise to the expressions of $R_1$ and $R_2$. We begin with the properties of $Q$ in the next section.


\setcounter{equation}{0}
\section{Properties of the collision operator and of its linearization}
\label{sec_propQ}


\subsection{Preliminaries}
\label{sub_prelim}

We recall the expressions of the gradient and divergence operator on the sphere. Let $x=(x_1,x_2,x_3)$ be a cartesian coordinate system associated with an orthonormal basis $(e_1,e_2,e_3)$ and let $(\theta, \phi)$ be a spherical coordinate system associated with this basis, i.e. $x_1= \sin \theta \cos \phi$, $x_2=  \sin \theta \sin \phi$, $x_3= \cos \theta$. Let also $(e_\theta, e_\phi)$ be the local basis associated with the spherical coordinate system ; the vectors  $e_\theta$ and $e_\phi$ have the following coordinates in the cartesian basis: $e_\theta = (\cos \theta \cos \phi, \cos \theta \sin \phi, -\sin \theta)$, $e_\phi = (- \sin \phi, \cos \phi, 0)$. Let $f(\omega)$ be a scalar function and $A= A_\theta e_\theta + A_\phi e_\phi$ be a tangent vector field. Then: 
$$\nabla_\omega f =  \partial_\theta f \,  e_\theta + \frac{1}{\sin \theta} \, \partial_\phi f \, e_\phi, \quad \nabla_\omega \cdot A = \frac{1}{\sin \theta} \partial_\theta (A_\theta \sin \theta) + \frac{1}{\sin \theta} \partial_\phi A_\phi. $$
$\Delta_\omega$ denotes the Laplace-Belltrami operator on the sphere: 
$$ \Delta_\omega f = \nabla_\omega \cdot \nabla_\omega f =   \frac{1}{\sin \theta} \partial_\theta ( \sin \theta \partial_\theta f )  + 
\frac{1}{\sin^2 \theta} \partial_{\phi \phi}  f  . $$

We write $F_{f^\varepsilon}$ for $F_0^\varepsilon$. We introduce the 'collision' operator, which corresponds to the leading order term of  (\ref{FP_mf_eps_0}): 
\begin{eqnarray} 
& & Q(f) = -  \nabla_\omega \cdot (F_f f) + d \Delta_\omega f ,  
\label{Q_def} \\
& & F_f = \nu \, \, (\mbox{Id} - \omega \otimes \omega) \Omega_f , 
\label{Force_def} \\
& & \Omega_f = \frac{j_f}{|\, \,  j_f\, \,  |}, \quad \mbox{ and } \quad j_f = \int_{ \omega \in {\mathbb S}^2 } 
\omega \, f \, d\omega
 \, . 
\label{Omega_def} 
\end{eqnarray}
We note that $Q(f)$ is a non linear operator. From now on, we assume that $f$ is as smooth and integrable as necessary. We note that $Q$ acts as an operator on functions of $\omega$ only and that the possible dependence of these functions on $(x,t)$ can be ignored. The properties of $Q$ have been demonstrated in Ref. \cite{Degond_M3AS08} are developed in the next section.


\subsection{Properties of $Q$}
\label{subsec_prop_Q}

\subsubsection{Null-space of $Q$}
\label{subsub_Q_null_space}

For $\Omega \in {\mathbb S}^2$, let $\mu = \cos \theta = (\omega \cdot \Omega)$. We denote by $\sigma(\mu)$ an antiderivative of $\nu(\mu)$, i.e. $(d \sigma / d \mu)(\mu) = \nu(\mu)$. We define 
\begin{eqnarray} 
& &  M_{\Omega}(\omega) = C \exp(\frac{\sigma(\omega \cdot \Omega)}{d}), \quad \int M_{\Omega}(\omega) \, d \omega =1 
 \, . 
\label{M_def} 
\end{eqnarray}
The constant $C$ is set by the normalization condition (second equality of (\ref{M_def}))~; it depends only on $d$ and on the function $\sigma$ but not on $\Omega$. We note that, when $\nu$ is constant, $\sigma(\cos \theta ) = \nu \, \cos \theta$ and that $M_{\Omega}$ is the so-called Von-Mises distribution. The Von-Mises distribution extends the notion of Gaussian for functions defined on the sphere and is also known as the circular Gaussian. In the present case, the Von-Mises distribution is centered at (or peaked at) $\Omega$.

The following lemma states what are the elements of the null-space of $Q$, i.e. what are the equilibria of the problem (see Ref. \cite{Degond_M3AS08} for the proof):
\begin{lemma}
(i) The operator $Q$ can be written as 
\begin{eqnarray} 
& &  Q(f) = d \, \, \nabla_\omega \cdot \left[ M_{\Omega_f} \nabla_\omega \left( \frac{f}{M_{\Omega_f}} \right) \right] ,    
\label{Q_self}
\end{eqnarray}
and we have 
\begin{eqnarray} 
& &  \hspace{-1cm} H(f) := \int_{\omega \in {\mathbb S}^2}  Q(f) \frac{f}{M_{\Omega_f}} \, d\omega = - d \, \, \int_{\omega \in {\mathbb S}^2} M_{\Omega_f} \left| \nabla_\omega \left( \frac{f}{M_{\Omega_f}} \right) \right|^2 \, d\omega \leq 0 .     
\label{Q_entrop}
\end{eqnarray}

\noindent
(ii) The equilibria, i.e. the functions $f(\omega)$ such that $Q(f) = 0$ form a three-dimensional manifold ${\mathcal E}$ given by 
\begin{eqnarray} 
& & {\mathcal E} = \{ \rho M_{\Omega}(\omega)\quad | \quad \rho \in {\mathbb R}_+, \quad \Omega \in {\mathbb S}^2
\} 
 \, , 
\label{equi_mani} 
\end{eqnarray}
and $\rho$ is the total mass while $\Omega$ is the direction of the flux of $\rho M_{\Omega}(\omega)$, i.e.
\begin{eqnarray} 
& &  \int_{\omega \in {\mathbb S}^2} \rho \, M_{\Omega}(\omega) \, d\omega = \rho, \label{Max_mass} \\
& & \Omega = \frac{j_{\rho M_{\Omega}}}{|\, j_{\rho M_{\Omega}}\, |}  \, , \quad j_{\rho M_{\Omega}} = \int_{\omega \in {\mathbb S}^2} \rho M_{\Omega}(\omega) \, \omega \,  d\omega   .   
\label{Max_flux}
\end{eqnarray}
Furthermore, $H(f) =0$ if and only if $f \in {\mathcal E}$. 
\label{lem_equi}
\end{lemma}

An elementary computation shows that the flux can be written
\begin{eqnarray} 
& &  j_{\rho M_{\Omega}} =  c_1 \rho \Omega , \quad c_1 = \langle \cos \theta \rangle_{M_\Omega}, \label{flux}
\end{eqnarray}
where for any function $g(\cos \theta)$, the symbol $\langle g(\cos \theta) \rangle_M$ denotes the average of $g$ over the probability distribution $M_\Omega$, i.e.
\begin{eqnarray} 
& &  \hspace{-1cm}
\langle g(\cos \theta) \rangle_M = \int M_\Omega(\omega) g(\omega \cdot \Omega) \, d\omega = \frac{\int_0^\pi g(\cos \theta) \exp ( \frac{\sigma(\cos \theta)}{d})  \, 
\sin \theta \, d \theta}{\int_0^\pi \exp ( \frac{\sigma(\cos \theta)}{d}) \,   \sin \theta \, d \theta}
. \label{brackets}
\end{eqnarray}
(\ref{flux}) defines the constant $c_1$ appearing in (\ref{mass_eq_0}). 

\subsubsection{Generalized collision invariants}
\label{subsub_GCI}

The second set of lemmas state what are the {\bf generalized} collision invariants of $f$. Indeed, we recall that the collision invariants are classically defined as the functions $\psi(\omega)$ such that 
\begin{eqnarray} 
\int_{\omega \in {\mathbb S}^2} Q(f) \, \psi \, d \omega = 0 , \quad \forall f . 
\label{Q_coll_invar}
\end{eqnarray}
However, it is readily seen that the linear vector space of collision invariants is of dimension one, while the hydrodynamic limit requires that its dimension be equal to the dimension of ${\mathcal E}$, which is $3$ in the present case. To find the missing collision invariants, we slightly weaken the definition. We fix $\Omega \in {\mathbb S}^2$ arbitrarily, and we define a Generalized Collision Invariant (or GCI) associated to $\Omega$ as a function $\psi$ which satisfies (\ref{Q_coll_invar}) only for functions $f$ with direction $\Omega_f = \Omega$. This constraint is linear and can be resolved by the introduction of a Lagrange multiplier. This leads to the following definition: 

\begin{definition}
Let $\Omega \in {\mathbb S}^2$ be given. $\psi(\omega)$ is a Generalized Collision Invariant (or GCI) associated to $\Omega$ if and only if 
\begin{eqnarray} 
\int_{\omega \in {\mathbb S}^2} Q(f) \, \psi \, d \omega = 0, \quad \forall f \quad \mbox{ such that } \quad \Omega_f = \Omega, 
\label{Q_coll_invar_2}
\end{eqnarray}
\label{def_gci}
\end{definition}

It is shown in Ref. \cite{Degond_M3AS08} that, using (\ref{Q_self}) and Green's formula, (\ref{Q_coll_invar_2}) leads to the following problem defining the GCI's associated to a direction $\Omega$: $\exists \beta \in {\mathbb R}^3$, such that $\beta \cdot \Omega = 0$ and 
\begin{eqnarray} 
\nabla_\omega \cdot ( M_{\Omega} \nabla_\omega \psi ) = \beta \cdot (\Omega \times \omega) M_{\Omega} . 
\label{Q_coll_invar_5}
\end{eqnarray}
This problem is obviously linear, so that the set ${\mathcal C}_\Omega$ of GCI's associated to $\Omega$ is a linear vector space. In a cartesian basis $(e_1, e_2, \Omega)$ and the associated spherical coordinates $(\theta, \phi)$, we have $\beta \cdot (\Omega \times \omega) = (- \beta_1 \sin \phi + \beta_2 \cos \phi) \sin \theta$ with $\beta_k = \beta \cdot e_k$, $k=1,2$.  Therefore, we can successively solve  for $\psi_1$ and $\psi_2$, the solutions of (\ref{Q_coll_invar_5}) with right-hand sides respectively equal to $- \sin \phi \sin \theta M_\Omega$ and $\cos \phi \sin \theta M_\Omega$. The following lemma provides the framework for solving (\ref{Q_coll_invar_5}). It is based on Lax-Milgram theorem and is proved in Ref. \cite{Degond_M3AS08}: 

\begin{lemma}
Let $\chi \in L^2({\mathbb S}^2)$ such that $\int \chi \, d\omega = 0$. The problem 
\begin{eqnarray} 
\nabla_\omega \cdot ( M_{\Omega} \nabla_\omega \psi ) = \chi , 
\label{CI_elliptic}
\end{eqnarray}
has a unique weak solution in the space ${\stackrel{\circ}{H^1}}({\mathbb S}^2)$, the quotient of the space $H^1({\mathbb S}^2)$ by the space spanned by the constant functions, endowed with the quotient norm. 
\label{lem_exist}
\end{lemma}

So, to each of the right-hand sides $\chi_1 = - \sin \phi \sin \theta M_\Omega$ or $\chi_2 = \cos \phi \sin \theta M_\Omega$ which have zero average on the sphere, there exist solutions  $\psi_1$ and $\psi_2$ respectively (unique up to constants) of problem (\ref{CI_elliptic}). We single out unique solutions by requesting that $\psi_1$ and $\psi_2$ have zero average on the sphere: $\int \psi_k \, d\omega = 0$, $k=1,2$. Then, we have, as a consequence of Lemma \ref{lem_exist}: 

\begin{proposition}
The set  ${\mathcal C}_\Omega$ of generalized collisional invariants associated with the vector $\Omega$ which belong to $H^1({\mathbb S}^2)$ is a three dimensional vector space ${\mathcal C}_\Omega = \mbox{Span} \{ 1, \psi_1, \psi_2 \}$.
\label{lem_coll_invar}
\end{proposition}

More explicit forms for $\psi_1$ and $\psi_2$ can be found. By expanding in Fourier series with respect to $\phi$, we easily see that 
\begin{eqnarray} 
& & \psi_1 = - g(\cos \theta) \sin \phi,  \quad \psi_2 = g(\cos \theta) \cos \phi , \label{psi_1_2} 
\end{eqnarray}
where $g(\mu)$ is a solution of the elliptic problem on $[-1,1]$:
\begin{eqnarray} 
& & - (1-{\mu}^2) \partial_\mu ( e^{\sigma(\mu)/d} (1-{\mu}^2) \partial_\mu g ) + e^{\sigma(\mu)/d} g = - (1-{\mu}^2)^{3/2} e^{\sigma(\mu)/d}
. \label{g} 
\end{eqnarray}
To solve this problem, we apply the following lemma:

\begin{lemma}
Let $X =  \{ g \, | \, (1-\mu^2)^{-1/2} g \in L^2(-1,1) \}$, \, $V = \{ g \in X \, | \, (1-\mu^2)^{1/2} \partial_\mu g \in L^2(-1,1) \}$. Let $\alpha (\mu)$ belong to $L^\infty(-1,1)$ such that there exists $\alpha_0>0$ and $\alpha(\mu) > \alpha_0$. Then, for any $f \in X$, there exists a unique solution $g \in V$ of the problem 
\begin{eqnarray} 
& & - (1-{\mu}^2) \partial_\mu ( e^{\sigma(\mu)/d} (1-{\mu}^2) \partial_\mu g ) + \alpha (\mu) g = f. \label{generic_1}
\end{eqnarray}
Additionally, the maximum principle holds: if $f$ is non-positive (respectively non-negative), then, so is $g$. 
\label{lem_ellipt_1D}
\end{lemma}

The next lemma, in the spirit of the previous one, will prove useful in the sequel: 
\begin{lemma}
Let ${\mathcal X} = L^2(-1,1)$,\, ${\mathcal V} = \{ g \in {\mathcal X} \, | \, (1-\mu^2)^{1/2} \partial_\mu g \in L^2(-1,1) \}$, \, $\stackrel{\circ}{\mathcal X} = \{ f \in {\mathcal X} \, | \, \int_{-1}^1 f \, d \mu = 0 \}$, \, $\stackrel{\circ}{\mathcal V} = {\mathcal V} / {\mathbb R}$. Then, for any $f \in \stackrel{\circ}{\mathcal X}$, there exists a unique solution $g \in \stackrel{\circ}{\mathcal V}$ of the problem 
\begin{eqnarray} 
& & - \partial_\mu ( e^{\sigma(\mu)/d} (1-{\mu}^2) \partial_\mu g ) = f.
\label{generic_2}
\end{eqnarray}
\label{lem_ellipt_1D_no0}
\end{lemma}

Both lemma are direct consequences of Lax-Milgram's theorem. The first one is proved in Ref.  \cite{Degond_M3AS08}. For the second Lemma, thanks to a  Poincar\'e inequality we note that the semi-norm of ${\mathcal V}$ is equivalent to the norm of ${\mathcal V}$ on the quotient $\stackrel{\circ}{\mathcal V}$. For both problems, no boundary conditions at $\pm 1$ need to be prescribed. This is due to the degeneracy of the elliptic operator at these points. 

Applying Lemma \ref{lem_ellipt_1D} shows that the function $g$, solution of (\ref{g}), is uniquely defined in the space $V$. For convenience, we introduce  $h(\mu) = (1-\mu^2)^{-1/2} g \in L^2(-1,1) $ or equivalently $h(\cos \theta) = g(\cos \theta) / \sin \theta$.
We then define 
\begin{eqnarray} 
& & \vec \psi_\Omega (\omega) = (\Omega \times \omega) \, h(\mu) = \psi_1 e_1 + \psi_2 e_2, \quad \mu = (\omega \cdot \Omega) \, 
. \label{psi} 
\end{eqnarray}
$\vec \psi_\Omega$ is the vector generalized collisional invariant associated to the direction $\Omega$ and is uniquely defined by the problem 
\begin{eqnarray} 
\nabla_\omega \cdot ( M_{\Omega} \nabla_\omega \vec \psi_\Omega ) =  (\Omega \times \omega) M_{\Omega} , \quad  \int_{{\mathbb S}^2} \vec \psi_\Omega \, d \omega  = 0 . 
\label{Prob_Psi}
\end{eqnarray}
We note that, by the maximum principle, $h \leq 0$.


\subsection{Linearization about the equilibrium state}
\label{sub_linearization}

We introduce a macro-micro decomposition of $f$:
\begin{equation}
f = \rho_f M_{\Omega_f} + \varphi, \label{micro_macro}
\end{equation}
$\rho_f$ and $\Omega_f$ being the density and mean velocity direction of $f$, i.e.  
\begin{equation}
\rho_f = \int_{{\mathbb S}^2} f \, d \omega, \label{rho_f}
\end{equation}
and $\Omega_f$ given by (\ref{Omega_def}). These definitions of $\rho_f$ and $\Omega_f$ are equivalent to saying that $\varphi$ belongs to the space: 
\begin{equation}
\Phi_{\Omega_f} = \left\{ \varphi \in L^1({\mathbb S}^2) \, \,  | \, \, \int_{{\mathbb S}^2} \varphi \, d \omega = 0 \, \mbox{ and } \, \Omega_f \times \int_{{\mathbb S}^2} \varphi \, \omega \, d \omega =0  \right\}. \label{space_phi}
\end{equation}
For any given $\Omega \in {\mathbb S}^2$, we define the following linear operator operating on $\Phi_\Omega$: 
\begin{eqnarray} 
& &  L_\Omega \varphi = d \, \, \nabla_\omega \cdot \left[ M_{\Omega} \nabla_\omega \left( \frac{\varphi}{M_{\Omega}} \right) \right].
\label{def_L}  
\end{eqnarray}
Inserting (\ref{micro_macro}) into (\ref{Q_self}) yields that 
\begin{eqnarray*} 
& &  Q(f) = L_{\Omega_f} (\varphi).
\end{eqnarray*}

We now precise the functional setting. Let ${\mathbb X}_\Omega = \{ \varphi \, | \, \int |\varphi|^2 \, M_\Omega^{-1} \, d \omega \, < \, \infty \}$, \, ${\mathbb V}_\Omega = \{ \varphi \in {\mathbb X}_\Omega \, | \, \int M_\Omega \, | \, \nabla_\omega ( M_\Omega^{-1} \, \varphi ) \, |^2  \, d \omega \, < \, \infty \}$, \, $\stackrel{\circ}{\mathbb X}_\Omega = {\mathbb X}_\Omega / (M_\Omega {\mathbb R})$, \, $\stackrel{\circ}{\mathbb V}_\Omega = {\mathbb V}_\Omega / (M_\Omega {\mathbb R})$. We perform the usual identification of ${\mathbb X}_\Omega$ with its dual ${\mathbb X}_\Omega'$. The following lemma states the properties of $L_{\Omega}$.  

\begin{lemma}
Consider $L_\Omega$ as an operator from ${\mathbb V}_\Omega $ into its dual ${\mathbb V}_\Omega'$, defined by the bilinear form on ${\mathbb V}_\Omega $: 
\begin{eqnarray} 
& &  \langle L_{\Omega} \varphi , \chi \rangle_{\langle {\mathbb V}_\Omega', {\mathbb V}_\Omega \rangle} = - d  \int_{{\mathbb S^2}} M_\Omega \nabla_\omega \, \left( \frac{\varphi}{M_\Omega} \right) \cdot \nabla_\omega \left( \frac{\chi}{M_\Omega} \right) \, d\omega. 
\label{L_bilinear}
\end{eqnarray}
Then,  

\noindent
(i) The null-space of $L_\Omega$ is the linear space spanned by $M_\Omega$. 

\noindent
(ii) Let $\zeta \in {\mathbb X}_\Omega$. The equation \, $ L_\Omega \varphi = \zeta $ \,  has a solution $\varphi$ in ${\mathbb V}_\Omega$  if and only if $\zeta$ satisfies the solvability condition: 
\begin{eqnarray} 
& &  \int_{{\mathbb S^2}}  \zeta \, d\omega = 0. 
\label{solv_L}
\end{eqnarray}
And $\varphi$ is unique if it additionally satisfies (\ref{solv_L}). This unique solution is written $\varphi = L_{\Omega}^{-1} \zeta$ and $L_{\Omega}^{-1}$ is called the pseudo-inverse of $L_{\Omega}$. 

\noindent
(iii) The solution $\varphi$ belongs to $\Phi_\Omega$ if and only if  $\zeta$ satisfies 
\begin{eqnarray} 
& &  \int_{{\mathbb S^2}}  \zeta \, \vec \psi_\Omega \, d\omega = 0, 
\label{add_L}
\end{eqnarray}
where $\vec \psi_\Omega$ is the GCI (\ref{psi}). 

\label{lem_prop_L}
\end{lemma}

\medskip
\noindent
{\bf Proof:} Statement (i) is obvious. To prove (ii), we note that the bilinear form (\ref{L_bilinear}) defines a bilinear form on the quotient space $\stackrel{\circ}{\mathbb V}_\Omega$. The solvability condition (\ref{solv_L}) is the necessary and sufficient condition for an element of ${\mathbb X}_\Omega$ to belong to the dual space $(\stackrel{\circ}{\mathbb X}_\Omega)'$. Then, the proof of (ii) follows from the application of Lax-Milgram theorem. Indeed, by Poincare's lemma, the bilinear form (\ref{L_bilinear}) defines a norm on $\stackrel{\circ}{\mathbb V}_\Omega $ which is equivalent to the norm of ${\mathbb V}_\Omega$. Therefore, there exists a unique solution $\varphi \in \stackrel{\circ}{\mathbb V}_\Omega$ of the problem 
\begin{eqnarray} 
& &  d \int_{{\mathbb S^2}} M_\Omega \nabla_\omega \, \left( \frac{\varphi}{M_\Omega} \right) \cdot \nabla_\omega \left( \frac{\chi}{M_\Omega} \right) \, d\omega = - \int_{{\mathbb S^2}} \zeta \, \chi  \, \frac{d\omega}{M_\Omega}, \label{var_formul}
\end{eqnarray}
which uniquely defines a solution in ${\mathbb V}_\Omega$ provided the cancellation condition (\ref{solv_L}) is imposed on $\varphi$. Finally, inserting $\chi = M_\Omega \vec \psi_\Omega$ in (\ref{var_formul}) and using (\ref{Prob_Psi}), we deduce that 
\begin{eqnarray*} 
& &  \int_{{\mathbb S^2}} \zeta \,  \vec \psi_\Omega \, d\omega = - d \int_{{\mathbb S^2}} M_\Omega \nabla_\omega \, \left( \frac{\varphi}{M_\Omega} \right) \cdot \nabla_\omega \vec \psi_\Omega \, d\omega = \\
& &  \hspace{2cm} =  d \int_{{\mathbb S^2}}   \, \frac{\varphi}{M_\Omega}  \nabla_\omega \cdot ( M_\Omega \nabla_\omega \vec \psi_\Omega) \, d\omega
= d \, \, \Omega \times \int_{{\mathbb S^2}}   \, \varphi  \omega   \, d\omega.
\end{eqnarray*}
Property (iii) is an immediate consequence of this identity. \endproof


\subsection{Coefficients of the hydrodynamic model}
\label{sub_coefs}

We finish these preliminaries by recalling the expressions of the coefficients $c_1$, $c_2$ and $c_3$ as they were derived in Ref. \cite{Degond_M3AS08}. For any functions $g(\cos \theta)$, $h(\cos \theta)$ with $h \geq0$, we denote by $\langle g \rangle_h$ the average of $g$ over the probability distribution defined by $h$:
\begin{eqnarray} 
& &  \langle g(\cos \theta) \rangle_h = \frac{\int g(\omega \cdot \Omega) \, h(\omega \cdot \Omega) \, d\omega}{\int h(\omega \cdot \Omega) \, d\omega} = \frac{\int_0^\pi g(\cos \theta)   \, h(\cos \theta)
\sin \theta \, d \theta}{\int_0^\pi h(\cos \theta) \,   \sin \theta \, d \theta}
. \label{brack_h}
\end{eqnarray}
Then, the constants $c_1$, $c_2$ and $c_3$ of model (\ref{mass_eq_0}), (\ref{Omega_eq_0}) are given by: 
\begin{eqnarray} 
& & \hspace{-1cm} c_1 = \langle \cos \theta \rangle_M \, , 
\quad  c_2 = \langle \cos \theta \rangle_{(\sin^2 \theta) \nu  h  M} \, , \quad c_3 = d \left\langle \frac{1}{\nu} \right\rangle_{(\sin^2 \theta) \nu  h  M}.
\label{const_hydro} 
\end{eqnarray}
With simple computations, one can check that these definitions are equivalently expressed by the following relations: 
\begin{eqnarray} 
& & \hspace{-1cm} \int_{{\mathbb S^2}} ((\omega \cdot \Omega) - c_1) \, M_\Omega \, d \omega = 0, \label{rel_c1} \\
& & \hspace{-1cm} \int_{{\mathbb S^2}} \frac{\nu}{d} \, ((\omega \cdot \Omega) - c_2) \, (1 - (\omega \cdot \Omega)^2) \,  h \, M_\Omega \, d \omega = 0, \label{rel_c2} \\
& & \hspace{-1cm}  \int_{{\mathbb S^2}} \left( 1 - \frac{\nu c_3}{d} \right) \, (1 - (\omega \cdot \Omega)^2) \,  h \, M_\Omega \, d \omega = 0. \label{rel_lambda} 
\end{eqnarray}


\setcounter{equation}{0}
\section{The Chapman-Enskog expansion}
\label{sec_CE}


\subsection{Setting up the expansion}
\label{subsec_CE_setting}

We introduce the macro-micro decomposition (\ref{micro_macro}) in a scaled form: 
\begin{equation}
f^{\varepsilon}= \rho^{\varepsilon} M_{\Omega^{\varepsilon}} + \varepsilon  G^{\varepsilon},
\label{mima_eps}
\end{equation}
where $\rho^{\varepsilon}= \rho_{f^{\varepsilon}}$ and $\Omega^{\varepsilon}= \Omega_{f^{\varepsilon}}$ are  the density and velocity direction of the solution $f^{\varepsilon}$ of the kinetic model (\ref{FP_mf_eps_0}) and where $G^\varepsilon \in \Phi_{\Omega^\varepsilon}$. This leads to: 
\begin{eqnarray} 
& & \hspace{-1cm}  (\partial_t + \omega \cdot \nabla ) (\rho^{\varepsilon} M_{\Omega^{\varepsilon}} + \varepsilon  G^{\varepsilon}) + \varepsilon \nabla_\omega \cdot (F_2^\varepsilon \rho^{\varepsilon} M_{\Omega^{\varepsilon}} ) = L_{\Omega_\varepsilon}(  G^{\varepsilon})  + O(\varepsilon^2) . 
\label{FP_mf_eps_1}
\end{eqnarray}
This justifies the scaling (\ref{mima_eps}) because $ G^{\varepsilon}$ appears as an $O(1)$ quantity, showing that the correction to equilibrium $\varepsilon  G^{\varepsilon}$ is $O(\varepsilon)$. 

Integrating (\ref{FP_mf_eps_1}) and using Lemma \ref{lem_prop_L} (ii), we find:
\begin{eqnarray} 
& & \hspace{-1cm}  \int_{{\mathbb S}^2} (\partial_t + \omega \cdot \nabla ) (\rho^{\varepsilon} M_{\Omega^{\varepsilon}} + \varepsilon  G^{\varepsilon}) \, d \omega = O(\varepsilon^2)  . 
\label{FP_int_rho}
\end{eqnarray}
The contribution of the term $\rho^{\varepsilon} M_{\Omega^{\varepsilon}}$ has been computed in Ref. \cite{Degond_M3AS08} and we get: 
\begin{eqnarray} 
& & \partial_t \rho^{\varepsilon} + \nabla \cdot (c_1 \rho^\varepsilon \Omega^{\varepsilon} ) =  \varepsilon R_1^{\varepsilon} + O(\varepsilon^2), \quad  \, R_1^{\varepsilon} = - \nabla \cdot \int_{{\mathbb S}^2}  G^{\varepsilon} \,  \omega \, d \omega  . \label{FP_rho} 
\end{eqnarray}
Indeed, since $G^\varepsilon \in \Phi_{\Omega^\varepsilon}$, we have
$$ \int_{{\mathbb S}^2}  \partial_t G^{\varepsilon} \, d \omega = \partial_t \left( \int_{{\mathbb S}^2}  G^{\varepsilon} \, d \omega \right) = 0.$$

Now, multiplying (\ref{FP_mf_eps_1}) by the GCI $\vec \psi_\Omega (\omega) = (\Omega \times \omega) \, h(\Omega \cdot \omega) $, integrating with respect to $\omega$ and using Lemma \ref{lem_prop_L} (iii) and that $G^\varepsilon \in \Phi_{\Omega^\varepsilon}$, we find:
\begin{eqnarray} 
& & \hspace{-1cm}  \Omega^\varepsilon \times \left\{ \int_{{\mathbb S}^2} \left[ (\partial_t + \omega \cdot \nabla ) (\rho^{\varepsilon} M_{\Omega^{\varepsilon}} + \varepsilon G^{\varepsilon}) + \varepsilon \nabla_\omega \cdot (F_2^\varepsilon \rho^{\varepsilon} M_{\Omega^{\varepsilon}} ) \right] \, h \, \omega \, d \omega \right\} = O(\varepsilon^2)  . 
\label{FP_int_Omega}
\end{eqnarray}
Now, thanks to the computations of Ref. \cite{Degond_M3AS08}, we have: 
\begin{eqnarray*} 
& & \hspace{-0.5cm} \Omega^\varepsilon \times \left\{ \int_{{\mathbb S}^2}  (\partial_t + \omega \cdot \nabla ) \rho^{\varepsilon} M_{\Omega^{\varepsilon}} \, h \, \omega \, d \omega \right\} = \frac{\langle \sin^2 \theta \nu h \rangle_{M_{\Omega^{\varepsilon}}}}{2d} \, \, \Omega^{\varepsilon} \times \left\{ \rho^{\varepsilon} \left( \partial_t \Omega^{\varepsilon} + \right. \right. \nonumber \\
& & \hspace{4cm}  \left. \left. + c_2 ( \Omega^{\varepsilon} \cdot \nabla ) \Omega^{\varepsilon} \right) +  c_3 ( \mbox{ Id } -  \Omega^{\varepsilon} \otimes \Omega^{\varepsilon} ) \nabla \rho^{\varepsilon} \right\} , 
 \end{eqnarray*}
from which we deduce that
\begin{eqnarray} 
& & \hspace{-1cm} \rho^{\varepsilon} \left( \partial_t \Omega^{\varepsilon} + c_2 ( \Omega^{\varepsilon} \cdot \nabla ) \Omega^{\varepsilon} \right) +  c_3 ( \mbox{ Id } -  \Omega^{\varepsilon} \otimes \Omega^{\varepsilon} ) \nabla \rho^{\varepsilon} = \varepsilon R_2 + O(\varepsilon^2) , \label{FP_Omega} \\
& & \hspace{-1cm} 
R_2 = - \frac{2d}{\langle \sin^2 \theta \nu h \rangle_{M_{\Omega^{\varepsilon}}}} \, (\mbox{Id} - \Omega^{\varepsilon}  \otimes \Omega^{\varepsilon}) \,  \int_{{\mathbb S}^2}  [ \, \partial_t G^{\varepsilon} +  \omega \cdot \nabla G^{\varepsilon} + \nonumber \\
& & \hspace{4cm} + \nabla_\omega \cdot (F_2^\varepsilon \rho^{\varepsilon} M_{\Omega^{\varepsilon}}) \, ] \, h  \, \omega   \, d \omega . \label{FP_Omega_2}
 \end{eqnarray}
 
If we omit the $O(\varepsilon^2)$ remainders in eqs (\ref{FP_rho}) and (\ref{FP_Omega_2}), we find a macroscopic model which approximates the moments of the Fokker-Planck model (\ref{FP_mf_eps_0}) up to $O(\varepsilon^2)$ terms. The goal is now to compute $R_1$ and $R_2$ and to show that, up to $O(\varepsilon)$ terms, they have the expressions given by Theorem \ref{theo_limit}. Obviously, this requires the computation of $G^\varepsilon$. 

From (\ref{FP_mf_eps_1}), we find that $G^\varepsilon =  \tilde G^\varepsilon + O(\varepsilon) $ where $\tilde G^\varepsilon$ is a solution of the problem
\begin{eqnarray} 
& & \hspace{-1cm}  
L_{\Omega_\varepsilon}(  \tilde G^{\varepsilon}) = 
(\partial_t + \omega \cdot \nabla ) (\rho^{\varepsilon} M_{\Omega^{\varepsilon}})  . 
\label{eq_tilde_G}
\end{eqnarray}
Therefore, replacing $G^\varepsilon$ by $\tilde G^\varepsilon$ in the expressions of $R_1$ and $R_2$ will not change the order of the approximation. From now on, we will omit the tildes and consider $G^\varepsilon$ as the solution of (\ref{eq_tilde_G}). 

Now, the plan is as follows 
\begin{enumerate}
\item Show that (\ref{eq_tilde_G}) has a unique solution $G^\varepsilon$ belonging to the space $\Phi_{\Omega^\varepsilon}$. This means proving that the right-hand side of (\ref{eq_tilde_G}) satisfies the solvability conditions (\ref{solv_L}) and (\ref{add_L}). 
\item Compute the expression of $G^\varepsilon$, i.e. invert (\ref{eq_tilde_G}).
\item Insert the expression of $G^\varepsilon$ in the definitions of $R_1$ and $R_2$ and compute them.
\end{enumerate}
We now successively perform these tasks. In the following, we will omit the exponents $\varepsilon$ to make the expressions lighter.


\subsection{Preliminary lemmas}
\label{subsec_prel_lem}

\medskip
We will have to deal with the integrals over $\omega$ of expressions involving tensor products of ${\mathcal O}_\bot$ and ${\mathcal O}_\parallel$. To this aim, we use a spherical coordinate system associated with a cartesian basis whose third basis vector coincides with $\Omega$. We denote by $(\theta,\phi)$ the associated angular coordinates as defined in section \ref{sub_prelim}. For a function $u(\omega)$, we define 
$$ ( u )_\phi = \frac{1}{2\pi} \int_0^{2 \pi} u(\theta,\phi) \, d \phi, $$
the average of $u$ over the angle $\phi$. The proof of the following lemma is easy and omitted:

\begin{lemma}
(i) For all odd tensor powers $p$,  we have 
\begin{eqnarray}
& & \hspace{-1cm}
( \, ({\mathcal O}_\bot \omega )^{\otimes p} \, )_\phi = 0 .
\label{odd_tens_pow}
\end{eqnarray} 

\noindent
(ii) The first even tensor powers of ${\mathcal O}_\bot \omega$ are given by:
\begin{eqnarray}
& & \hspace{-1cm}
( \, ({\mathcal O}_\bot \omega )^{\otimes 2} \, )_\phi = \frac{1}{2} \sin^2 \theta \, {\mathcal O}_\bot ,
\label{square_tens_pow} \\
& & \hspace{-1cm}
( \, ({\mathcal O}_\bot \omega )^{\otimes 4} \, )_\phi = \frac{1}{8} \sin^4 \theta \, {\mathbb O}_\bot ,
\label{fourth_tens_pow}
\end{eqnarray} 
where we define the fourth order tensor ${\mathbb O}_\bot$:
\begin{eqnarray}
& & \hspace{-1cm}
{\mathbb O}_\bot = ({\mathcal O}_\bot)^{\otimes 2} + \left( ({\mathcal O}_\bot)^{\otimes 4} \right)_{[26],[48]} + \left( ({\mathcal O}_\bot)^{\otimes 4} \right)_{[28],[46]}
,
\label{def_big_O}
\end{eqnarray} 
and the subscript $[ij]$ denotes contraction with respect to indices $i$ and $j$. Restricted to the plane normal to $\Omega$, ${\mathbb O}_\bot$ can be written:
\begin{eqnarray} 
({\mathbb O}_\bot)_{ijkl} = \delta_{ij}\delta_{kl} + \delta_{ik} \delta_{jl} + \delta_{il} \delta_{jk} , \label{express_big_O}
\end{eqnarray} 
where $(\delta_{ij})_{i,j \in \{1,2\}}$ is the two-dimensional Kronecker tensor. We note that ${\mathbb O}_\bot$ is invariant under all rotations of the plane normal to $\Omega$, i.e. it satisfies
$$ ({\mathbb O}_\bot)_{ijkl} R_{ii'} R_{jj'} R_{kk'} R_{ll'}= ({\mathbb O}_\bot)_{i'j'k'l'}, $$ 
for all rotations $R_{ii'}$ of the plane normal to $\Omega$, where Einstein's repeated index summation rule is assumed. 
\label{lem_phi_averages}
\end{lemma}


\subsection{Sovability of the equation for $G$}
\label{subsec_solvability}

We state the first lemma: 

\begin{lemma}
We have: 
\begin{eqnarray} 
(\partial_t + \omega \cdot \nabla ) (\rho M_{\Omega})  & = &A_\parallel \cdot \nabla_\parallel \rho + A_\bot \cdot \nabla_\bot \rho \nonumber\\
&+&\rho (B_{\bot, \bot} : \nabla_{\bot, \bot} \Omega + B_{\parallel, \bot} : \nabla_{\parallel, \bot} \Omega) + O(\varepsilon),  
\label{eq_pt+omnax_rhoMom}
\end{eqnarray}
where
\begin{eqnarray} 
& & \hspace{-1cm}   
A_\bot = M_{\Omega} \left( 1 - \frac{\nu c_3}{d} \right) {\mathcal O}_\bot \omega \, , \label{Aperp} \\
& & \hspace{-1cm}   
A_\parallel = M_{\Omega} ( (\omega \cdot \Omega) - c_1) \Omega \, , \label{Apar} \\
& & \hspace{-1cm}   
B_{\bot, \bot} = M_{\Omega} \left( \frac{\nu}{d} ({\mathcal O}_\bot \omega) \otimes ({\mathcal O}_\bot \omega) - c_1 {\mathcal O}_\bot \right) \, , \label{Bperp} \\
& & \hspace{-1cm}   
B_{\parallel, \bot} = M_{\Omega} \frac{\nu}{d} (
(\omega \cdot \Omega) - c_2) \Omega \otimes ({\mathcal O}_\bot \omega) \, , \label{Bparperp} 
\end{eqnarray}
and where `:' denotes the contracted product of two tensors. 
\label{lem_pt+omnax_rhoMom}
\end{lemma}

{\bf Proof:} We have 
\begin{eqnarray*} 
& & \hspace{-1cm}  
(\partial_t + \omega \cdot \nabla ) (\rho M_{\Omega})   = M_{\Omega} \left( 
(\partial_t + \omega \cdot \nabla ) \rho + \rho  \frac{\partial (\ln M_\Omega )}{\partial \Omega} (\partial_t + \omega \cdot \nabla ) \Omega
\right).
\end{eqnarray*}
Classically, in the Chapman-Enskog procedure, time derivatives are replaced by space derivatives, using the following identities
\begin{eqnarray*} 
& & \partial_t \rho = -  \nabla \cdot (c_1 \rho \Omega )  + O(\varepsilon) , \\
& & \hspace{-1cm} \rho \partial_t \Omega = -  c_2 \rho ( \Omega \cdot \nabla ) \Omega -  c_3 \, {\mathcal O}_\bot   \nabla \rho + O(\varepsilon),  \end{eqnarray*}
which are deduced from (\ref{FP_rho}) and (\ref{FP_Omega}). For any tangent vector $\dot \Omega$ to ${\mathbb S}^2$ at $\Omega$, we have:
\begin{eqnarray*} 
& & \frac{\partial (\ln M_\Omega )}{\partial \Omega} \dot \Omega = \frac{\nu}{d} (\omega \cdot \dot \Omega) .
\end{eqnarray*}
Then, we note that: 
\begin{eqnarray} 
& & \omega \cdot \nabla \rho = {\mathcal O}_\bot \omega \cdot \nabla_\bot \rho + (\omega \cdot \Omega) \Omega \cdot \nabla_\parallel \rho, \nonumber \\
& & \nabla \cdot \Omega = {\mathcal O}_\bot : (\nabla_{\bot, \bot} \Omega),\nonumber\\
& & \omega \cdot (\Omega \cdot \nabla) \Omega = (\Omega \otimes {\mathcal O}_\bot \omega) : \nabla_{\parallel,\bot} \Omega, \nonumber \\
& & \omega \cdot (\omega \cdot \nabla) \Omega = ({\mathcal O}_\bot \omega \otimes {\mathcal O}_\bot \omega ): \nabla_{\bot,\bot} \Omega + (\omega \cdot \Omega) \, (\Omega \otimes {\mathcal O}_\bot \omega) : \nabla_{\parallel,\bot} \Omega.  \label{div_Omega_2}
\end{eqnarray}
Collecting these identities, we find expressions (\ref{Aperp}) to (\ref{Bparperp}). 
\endproof

\begin{lemma}
The quantities 
$A_\bot$, $A_\parallel$ $B_{\bot,\bot}$, $B_{\parallel, \bot}$ satisfy (separately) conditions 
(\ref{solv_L}) and (\ref{add_L}). As vectors or tensors, this means that they satisfy these conditions componentwise. 
\label{lem_solv_AB}
\end{lemma}

{\bf Proof:} We summarize the main arguments and leave the computational details to the reader. 

\noindent
(i) $A_\bot$ satisfies (\ref{solv_L}) because of (\ref{odd_tens_pow}) and (\ref{add_L}) as a consequence of (\ref{rel_lambda}). 

\noindent
(ii) $A_\parallel$ satisfies (\ref{solv_L}) as a consequence of (\ref{rel_c1}) and (\ref{add_L}) because of (\ref{odd_tens_pow}). 

\noindent
(iii) $B_{\bot,\bot}$ satisfies (\ref{solv_L}) as a consequence of (\ref{rel_c1}) (after an integration by parts with respect to $\theta$) and (\ref{add_L}) because of (\ref{odd_tens_pow}).

\noindent 
(iv) $B_{\parallel, \bot}$ satisfies (\ref{solv_L}) because of (\ref{odd_tens_pow}) and (\ref{add_L}) as a consequence of (\ref{rel_c2}). \endproof


\subsection{Computation of $G$}
\label{subsec_geps_comput}

We have shown that the right-hand side of (\ref{eq_tilde_G}) can be decomposed into four different terms, corresponding to derivatives of $\rho$ and $\Omega$ in the directions normal or parallel to $\Omega$, and that each of these four components satisfies the solvability conditions (\ref{solv_L}) and (\ref{add_L}) separately. We now compute the pseudo inverse $L_{\Omega}^{-1}$ applied to these four components. 

\begin{lemma}
We have: 
\begin{eqnarray} 
& & \hspace{-1cm}   
\tilde A_\bot := - L_{\Omega}^{-1} A_\bot =  M_{\Omega} \, a_\bot \, {\mathcal O}_\bot \omega \, , \label{L-1_Aperp} \\
& & \hspace{-1cm}   
\tilde A_\parallel := - L_{\Omega}^{-1} A_\parallel = M_{\Omega} \,  a_\parallel  \, \Omega  \, , \label{L-1_Apar} \\
& & \hspace{-1cm}   
\tilde B_{\bot,\bot} := - L_{\Omega}^{-1} B_{\bot,\bot} = M_{\Omega} \left\{ b_1 \, ({\mathcal O}_\bot \omega) \otimes ({\mathcal O}_\bot \omega) \, + \,  b_2 \, {\mathcal O}_\bot \right\}  \, , \label{L-1_Bperp} \\
& & \hspace{-1cm}   
\tilde B_{\parallel, \bot} := - L_{\Omega}^{-1} B_{\parallel, \bot} =  M_{\Omega}  \, b_\parallel \, \, \Omega \otimes ({\mathcal O}_\bot \omega) \, , \label{L-1_Bparperp} 
\end{eqnarray}
where $a_\bot$, $a_\parallel$, $b_1$, $b_2$ and $b_\parallel$ are functions of $\omega \cdot \Omega$. They  are defined by the following relations (letting $\mu = \omega \cdot \Omega$):
\begin{enumerate}
\item  $\tilde a_\bot = a_\bot (\mu) \sqrt{1 - \mu^2}$ is the unique solution of (\ref{generic_1}) with 
\begin{eqnarray} 
& & \hspace{-1cm}  \alpha (\mu) = e^{\frac{\sigma}{d}} , \quad f(\mu) = \frac{1}{d}  e^{\frac{\sigma}{d}} \left( 1 - \frac{c_3 \nu(\mu)}{d} \right) (1 - \mu^2)^{3/2} .
\label{abot_rhs}
\end{eqnarray}
\item  $a_\parallel$ is the unique (up to an additive constant) solution of (\ref{generic_2}) with \begin{eqnarray} 
& & \hspace{-1cm}  f(\mu) = \frac{1}{d}  e^{\frac{\sigma}{d}} (\mu - c_1) , \label{apar_rhs}
\end{eqnarray}
and the constant is adjusted in such a way that $\int_{{\mathbb S}^2} \tilde A_\parallel \, d \omega= 0$. 
\item $\tilde b_1 = b_1  (1-\mu^2)$ is the unique solution of (\ref{generic_1}) with 
\begin{eqnarray} 
& & \hspace{-1cm}  \alpha (\mu) = 4 e^{\frac{\sigma}{d}} , \quad f(\mu) = \frac{\nu}{d^2}  e^{\frac{\sigma}{d}}  (1 - \mu^2)^2.
\label{b1_rhs}
\end{eqnarray}
\item $b_2$ is the unique (up to an additive constant) solution of (\ref{generic_2}) with 
\begin{eqnarray} 
& & \hspace{-1cm}  f(\mu) = e^{\frac{\sigma}{d}} \, ( 2 b_1 - \frac{c_1}{d}) , \label{b2_rhs}
\end{eqnarray}
and the constant is adjusted in such a way that 
$\int_{{\mathbb S}^2} \tilde B_\bot \, d \omega= 0$.  
\item $\tilde b_\parallel = b_\parallel \sqrt{1 - \mu^2} $ is the unique solution of (\ref{generic_1}) with 
\begin{eqnarray} 
& & \hspace{-1cm}  \alpha (\mu) = e^{\frac{\sigma}{d}} , \quad f(\mu) = \frac{\nu}{d^2}  e^{\frac{\sigma}{d}} (\mu - c_2) (1 - \mu^2)^{3/2} . 
\label{bparper_rhs}
\end{eqnarray}
\end{enumerate}
\label{lem_L-1}
\end{lemma}

\medskip
\noindent
{\bf Proof:} Preliminaries:  using spherical coordinates, we check that if $\varphi$ is of the form $\varphi = M_{\Omega} \, C_k(\cos \theta) \, \cos k \phi$, then, 
\begin{eqnarray} 
& & \hspace{-1cm}  
- \frac{1}{M_\Omega} L_\Omega \varphi 
= \frac{d e^{- \sigma /d}}{1-\mu^2} \, \cos k \phi \, \left[ - (1-\mu^2) \partial_\mu ( e^{ \sigma /d} (1-\mu^2) \partial_\mu C_k) + k^2 e^{ \sigma /d} C_k \right] , 
\label{Fourier_decomp}
\end{eqnarray}
with $\mu = \cos \theta$.  Similarly, if $\varphi = M_{\Omega} \, S_k(\cos \theta) \, \sin k \phi$, then $S_k$ satisfies the same identity with $\cos k \phi$ replaced by $\sin k \phi$.

Proof of (i): It is a matter of computation to show that $\tilde A_\bot$ defined by (\ref{L-1_Aperp}) is a solution of $- L_{\Omega} \tilde A_\bot = A_\bot$ provided that $\tilde a_\bot$ satisfies (\ref{generic_1}) with data $\alpha$ and $f$ given by (\ref{abot_rhs}) (use (\ref{Fourier_decomp}) with $k=1$). Now, it is clear that $\tilde A_\bot$ satisfies the normalization condition (\ref{solv_L}), because of (\ref{odd_tens_pow}). Therefore, $\tilde A_\bot$ is the unique solution called $- L_{\Omega}^{-1} A_\bot$. 

\noindent (ii) Using (\ref{Fourier_decomp}) with $k=0$, we show that $\tilde A_\parallel$ defined by (\ref{L-1_Apar}) is a solution of $- L_{\Omega} \tilde A_\parallel = A_\parallel$ provided that $\tilde a_\parallel$ satisfies (\ref{generic_2}) with data $f$ given by (\ref{apar_rhs}). $\tilde a_\parallel$ is defined up to an additive constant, which means that $\tilde A_\parallel$ is defined up to the addition of a function proportional to $M_{\Omega} \Omega$. The coefficient can be chosen in such a way that condition (\ref{solv_L}) is satisfied. This solution is the unique solution called $- L_{\Omega}^{-1} A_\parallel$. 

\noindent (iii) We proceed similarly. Using (\ref{Fourier_decomp}) successively with $k=2$ and $k=0$ we show that $- L_{\Omega} \tilde B_{\bot,\bot} = B_{\bot,\bot}$ provided that $b_1$ and $b_2$ are specified as stated in the theorem. Additionally, with (\ref{square_tens_pow}), we have 
$$ (\tilde B_{\bot,\bot})_\phi = (b_1 \frac{\sin^2 \theta}{2} + b_2) {\mathcal O}_\bot , $$
and since $b_2$ is defined up to a constant, we can adjust this constant to satisfy the normalization condition (\ref{solv_L}). The so-defined $\tilde B_{\bot,\bot}$ is the unique $- L_{\Omega}^{-1} B_{\bot,\bot}$. Note that, because of the factor $1-\mu^2$ in the expression of $f$ in (\ref{b1_rhs}), it is an easy matter to show that $b_1 = \tilde b_1 /(1-\mu^2)$ belongs to $L^2$ and that the assumptions for the application of lemma \ref{lem_ellipt_1D_no0} are satisfied. 

\noindent (iv) We proceed exactly in the same way for $\tilde B_{\parallel, \bot}$. Using (\ref{Fourier_decomp}) with $k=1$, we find that $\tilde B_{\parallel,\bot}$ is a solution of 
$- L_{\Omega} \tilde B_{\parallel,\bot} = B_{\parallel,\bot}$ provided that $\tilde b_\parallel$ satisfies (\ref{bparper_rhs}). The normalization condition (\ref{solv_L}) is satisfied because of (\ref{odd_tens_pow}) which proves that the so-defined $\tilde B_{\parallel,\bot}$ is the unique $- L_{\Omega}^{-1} B_{\parallel,\bot}$.
\endproof

\begin{lemma}
The following relations are satisfied: 
\begin{eqnarray}
& & \hspace{-1cm}  
\langle a_\bot \sin^2 \theta \rangle_{M_{\Omega}}  = 0, \quad  
\langle a_\parallel \rangle_{M_{\Omega}} = 0 
, \label{int_a} \\
& & \hspace{-1cm}  
\langle \frac{1}{2}  b_1 \sin^2 \theta +  b_2  \rangle_{M_{\Omega}}= 0 , \quad 
\langle b_\parallel \sin^2 \theta \rangle_{M_{\Omega}}  = 0 . 
\label{int_b}
\end{eqnarray}
\label{lem_rel_coef_tilde}
\end{lemma}
\medskip
\noindent
{\bf Proof:} Since $\tilde A_\bot$, $\tilde A_\parallel$, \ldots belong to the space $\Phi_{\Omega}$, their integral against $1$ and $\Omega \times \omega$ over $\omega \in {\mathbb S}^2$ vanishes (see definition (\ref{space_phi})). Using lemma \ref{lem_phi_averages}, this leads to the above listed relations. \endproof

Finally, as a consequence of lemma \ref{lem_L-1}, we can summarize: 

\begin{lemma} we have:  
\begin{eqnarray} 
&&- G = - L_{\Omega}^{-1} ((\partial_t + \omega \cdot \nabla ) (\rho M_{\Omega})) \nonumber \\
&=& \tilde A_\parallel \cdot \nabla_\parallel \rho + \tilde A_\bot \cdot \nabla_\bot \rho +\rho ( \tilde B_{\bot, \bot} : \nabla_{\bot, \bot} \Omega + \tilde B_{\parallel, \bot} : \nabla_{\parallel, \bot} \Omega) + O(\varepsilon) \, .
\label{express_Geps}
\end{eqnarray}
We can decompose $G$ into even and odd powers of ${\mathcal O}_\bot \omega$ and write
\begin{eqnarray} 
& & \hspace{-1cm} 
G =  G_{e} + G_{o} + O(\varepsilon), \label{Geps_odd_even} \\
& & \hspace{-1cm} 
- G_{e} = \tilde A_\parallel \cdot \nabla_\parallel \rho + \rho \tilde B_{\bot, \bot} : \nabla_{\bot, \bot} \Omega \nonumber \\
& & \hspace{-0.1cm} 
= M_{\Omega} \, \left\{  a_\parallel \, \Omega \cdot \nabla \rho +  \rho \left[ b_1 \, (({\mathcal O}_\bot \omega) \otimes ({\mathcal O}_\bot \omega)) : \nabla \Omega \, +\, b_2 \, {\mathcal O}_\bot : \nabla \Omega  \right] \right\} , \label{Geps_even} \\
& & \hspace{-1cm} 
- G_{o} = \tilde A_\bot \cdot \nabla_\bot \rho +\rho \tilde B_{\parallel, \bot} : \nabla_{\parallel, \bot} \Omega \nonumber\\ 
& & \hspace{-0.1cm} 
=M_{\Omega} \, \left\{ a_\bot \, {\mathcal O}_\bot \omega \cdot \nabla \rho +  \rho  b_\parallel \, (\Omega \otimes ({\mathcal O}_\bot \omega)) : \nabla \Omega \right\}.
\label{Geps_odd}
\end{eqnarray}
\label{lem_Geps}
\end{lemma}

\noindent
In the sequel, we will omit to mention the $O(\varepsilon)$ remainder. It should be understood that all results are up to a term of this order.


\subsection{Computation of the $O(\varepsilon)$ corrections}
\label{subsec_rhs}

\subsubsection{Computation of $R_1$ (\ref{FP_rho})}
\label{subsubsec_rhs_mass} 

In this section, we compute $R_1$, the right-hand side of (\ref{FP_rho}). Its expression is given in the following statement:  

\begin{lemma}
$R_1$ is given by formula (\ref{express_rhs_FP_rho}) with 
\begin{eqnarray}
& & \hspace{-1cm}  
\beta  = \langle a_\parallel \cos \theta \rangle_{M_{\Omega}} 
\, , \quad   
\gamma =  \langle (\frac{1}{2} b_1 \sin^2 \theta + b_2)  \cos \theta \rangle_{M_{\Omega}}
\, . 
\label{al_1_2}
\end{eqnarray}
\label{lem_rhs_FP_rho}
\end{lemma} 

Note: compared with (\ref{int_a}) and (\ref{int_b}), there is an additional factor $\cos \theta$ inside the brackets. 

\medskip
\noindent
{\bf Proof:} We multiply (\ref{express_Geps}) by $\omega$ and integrate over $\omega$. But, because $G \in \Phi_{\Omega}$, the normal component of $\int_{{\mathbb S}^2}  G \,  \omega \, d \omega$ to $\Omega$ vanishes and the projection upon $\Omega$ is the only non-zero component. It is obtained by multiplying (\ref{express_Geps}) by $(\Omega \cdot \omega) \Omega$ and integrating upon $\omega$. In this integration, the contribution of the odd part $G_{o}$ vanishes by (\ref{odd_tens_pow}). The contribution of the even part $G_{e}$ is readily found to be $\beta (\Omega \cdot \nabla_\parallel \rho) \Omega + \gamma \rho ({\mathcal O}_\bot : (\nabla \Omega)_{\bot,\bot}) \, \Omega$. The proof is ended by using (\ref{div_Omega_2}). 
\endproof

\begin{lemma}
We have \, $\beta  >0$. 
\label{lem_al1>0}
\end{lemma}

\medskip
\noindent
{\bf Proof:} We can write: 
\begin{eqnarray*}
\beta &=& \int_{{\mathbb S}^2} a_\parallel \, M_{\Omega} \, (\omega \cdot \Omega) \, d \omega 
= \int_{{\mathbb S}^2} a_\parallel \, M_{\Omega} \, ((\omega \cdot \Omega) - c_1) \, d \omega \\ 
&=& \int_{{\mathbb S}^2} (a_\parallel \, M_{\Omega}) \, (((\omega \cdot \Omega) - c_1) \, M_{\Omega}) \, M_{\Omega}^{-1} \, d \omega \\ 
&=& - \int_{{\mathbb S}^2} (a_\parallel \, M_{\Omega}) \, L_{\Omega} (a_\parallel \, M_{\Omega}) \, M_{\Omega}^{-1} \, d \omega\geq  0, 
\end{eqnarray*}
by the non-positivity of $L_{\Omega} $ (see (\ref{L_bilinear})). In the second equality, we have used that $a_\parallel \, M_{\Omega}$ satisfies (\ref{solv_L}) (see also (\ref{int_a})). The third equality is obvious and the fourth one is just using the definition of $a_\parallel \, M_{\Omega}$ (see (\ref{L-1_Apar})). $\beta$ is strictly positive, otherwise, $a_\parallel \, M_{\Omega}$ would belong to the kernel which of $L_{\Omega}$, which is spanned by $M_{\Omega}$. Since, besides that, $a_\parallel \, M_{\Omega}$ satisfies (\ref{solv_L}), it would be identically zero. But, applying $L_{\Omega}$ to it, then $((\omega \cdot \Omega) - c_1) \, M_{\Omega}$ would also be identically zero, which is obviously not the case. This concludes the proof by contradition. \endproof

\subsubsection{Computation of $R_2$ (\ref{FP_Omega_2})}
\label{subsubsec_rhs_Omega}

\begin{lemma}
$R_2$ is given by (\ref{express_rhs_FP_Omega}). The coefficients are given by 
(\ref{expr_mathcal_Q_1}), (\ref{expr_mathcal_Q_2}), (\ref{expr_mathcal_D}). 
\label{lem_rhs_FP_Omega}
\end{lemma} 

\medskip
\noindent
{\bf Proof:} We first compute the term involving $\partial_t G$:
\begin{eqnarray*}
T = - {\mathcal O}_\bot \,  \int_{{\mathbb S}^2}   \partial_t G \, h \, \omega   \, d \omega .
\end{eqnarray*}
We have 
\begin{eqnarray*}
T  &=&  - {\mathcal O}_\bot \, \partial_t \left( \int_{{\mathbb S}^2} G \, h \, \omega   \, d \omega \right) + {\mathcal O}_\bot \,  \left( \int_{{\mathbb S}^2}    G \, h' \, \omega \otimes \omega   \, d \omega \right) \,   \partial_t \Omega \, = \, T_1 + T_2, 
\end{eqnarray*}
where $h'$ denotes the derivative of $h$ with respect to $\mu = \omega \cdot \Omega$. For $T_1$, we decompose $\omega$ according to transverse and normal components:
\begin{eqnarray*}
T_1 =   - {\mathcal O}_\bot \, \partial_t \left( \int_{{\mathbb S}^2} G_o \, h \,  {\mathcal O}_\bot\omega   \, d \omega \right) - {\mathcal O}_\bot \, \partial_t \left( \int_{{\mathbb S}^2} G_e \, h \, \, (\omega \cdot \Omega )  \, \Omega  \, d \omega \right) = T_1^1 + T_1^2 , 
\end{eqnarray*}
where we have used (\ref{Geps_odd_even}) and (\ref{odd_tens_pow}) to introduce the even and odd parts of $G$. Thanks to (\ref{square_tens_pow}) and (\ref{div_Omega_2}), we find
\begin{eqnarray*}
& & \hspace{-1cm}
T_1^1 =   \lambda^1_{11} \,  {\mathcal O}_\bot \, \partial_t ({\mathcal O}_\bot \nabla \rho) + \lambda^1_{12} \, {\mathcal O}_\bot \, \partial_t (\rho (\Omega \cdot \nabla) \Omega),  \\
& & \hspace{-1cm}
T_1^2 =   \lambda^2_{11} \,  {\mathcal O}_\bot \,  \partial_t ((\Omega \cdot \nabla \rho) \Omega)
+ \lambda^2_{12} \, {\mathcal O}_\bot \, \partial_t ( \rho ( \nabla \cdot \Omega) \Omega ),  
\end{eqnarray*}
with 
\begin{eqnarray*}
& & \hspace{-1cm}
\lambda^1_{11} = \langle \frac{1}{2} \sin^2 \theta \, a_\bot \, h \rangle_{M_{\Omega}} ,  \quad 
\lambda^1_{12} = \langle \frac{1}{2} \sin^2 \theta \, b_\parallel \, h \rangle_{M_{\Omega}},  \\
& & \hspace{-1cm}
\lambda^2_{11} =   \langle \cos \theta \, a_\parallel \, h \rangle_{M_{\Omega}} , \quad 
\lambda^2_{12} = \langle (\frac{1}{2} \sin^2 \theta \cos \theta \, b_1 + \cos \theta \, b_2) \, h  \rangle_{M_{\Omega}} .  
\end{eqnarray*}
We proceed similarly for $T_2$. Since $ \omega \cdot \partial_t \Omega = ({\mathcal O}_\bot \omega) \cdot \partial_t \Omega$, we find:
\begin{eqnarray*}
T_2 & = &
\left( \int_{{\mathbb S}^2}    G_e \, h' \, ({\mathcal O}_\bot \omega \otimes {\mathcal O}_\bot \omega )  \, d \omega \right) \partial_t \Omega . 
\end{eqnarray*}
And thanks to (\ref{square_tens_pow}), (\ref{fourth_tens_pow}) and (\ref{express_big_O}), we find 
\begin{eqnarray}
& & \hspace{-1cm}
T_2 =   - \lambda_{21} \,  (\Omega \cdot \nabla \rho) \, \partial_t \Omega - \lambda_{22} \,  \rho ( \nabla \cdot \Omega) \partial_t \Omega - \lambda_{23} \, \rho \,  \sigma(\Omega) \partial_t \Omega, \label{T21}
\end{eqnarray}
with 
\begin{eqnarray*}
& & \hspace{-1cm}
\lambda_{21} =  \langle \frac{1}{2} \sin^2 \theta \, a_\parallel \, h' \rangle_{M_{\Omega}}, \quad 
\lambda_{22} =  \langle ( \frac{1}{4} \sin^4 \theta \, b_1 + \frac{1}{2} \sin^2 \theta \, b_2) \, h' \rangle_{M_{\Omega}}  , \nonumber \\
& & \hspace{-1cm}
\lambda_{23} = \langle \frac{1}{8} \sin^4 \theta \, b_1 \, h' \rangle_{M_{\Omega}} ,
\end{eqnarray*}
where we have used that 
\begin{eqnarray}
({\mathbb O}_\bot)_{ijkl} \partial_k \Omega_l &=& (\nabla_{\bot, \bot} \Omega + (\nabla_{\bot, \bot} \Omega)^T + (\nabla \cdot \Omega) {\mathcal O}_\bot)_{ij} \nonumber \\
&=& (\sigma(\Omega) + 2 (\nabla \cdot \Omega) {\mathcal O}_\bot)_{ij} \, \, . \label{BigO_bot_nabla_Omega}
\end{eqnarray}
Now, we note the following relations: 
\begin{eqnarray*}
& & \hspace{-1cm}
{\mathcal O}_\bot (\partial_t {\mathcal O}_\bot \nabla \rho ) = {\mathcal O}_\bot  \partial_t \nabla \rho - (\Omega \cdot \nabla \rho ) \partial_t \Omega \, ,  \\
& & \hspace{-1cm}
{\mathcal O}_\bot (\partial_t ( \rho (\Omega \cdot \nabla) \Omega)) = \partial_t \rho  (\Omega \cdot \nabla) \Omega + \rho (\partial_t \Omega \cdot \nabla) \Omega + \rho {\mathcal O}_\bot (\Omega \cdot \nabla) \partial_t \Omega  \, , \\
& & \hspace{-1cm}
(\partial_t \Omega \cdot \nabla) \Omega =  (\nabla \Omega)^T \partial_t \Omega = (\nabla \Omega)_{\bot,\bot}^T \partial_t \Omega 
\\
& & \hspace{2cm} =
\frac{1}{2} (\sigma(\Omega) - \Gamma(\Omega)) \partial_t \Omega + \frac{1}{2} (\nabla \cdot \Omega) \partial_t \Omega   \, , \\
& & \hspace{-1cm}
{\mathcal O}_\bot \partial_t ((\Omega \cdot \nabla \rho ) \Omega) = (\Omega \cdot \nabla \rho ) \partial_t \Omega  \, , \\
& & \hspace{-1cm}
{\mathcal O}_\bot \partial_t ( \rho (\nabla \cdot \Omega ) \Omega) = \rho (\nabla \cdot \Omega ) \partial_t \Omega  \, . 
\end{eqnarray*}
Collecting all these identities, we get: 
\begin{eqnarray}
& & \hspace{-1cm}
T = \lambda'_1 \, {\mathcal O}_\bot \, \nabla 
\partial_t\rho 
\, + \, \lambda'_2  \, (\Omega \cdot \nabla \rho) \,  \partial_t \Omega 
\, + \, \lambda'_3 \, ((\Omega \cdot \nabla) \Omega) \, \partial_t \rho 
\, + \, \lambda'_4  \, \rho \, \sigma (\Omega) \partial_t \Omega  \nonumber \\
& & \hspace{-0.4cm}
\, + \, \lambda'_5 \, \rho \, \Gamma (\Omega)  \, \partial_t \Omega  
\, + \, \lambda'_6 \, \rho \, (\nabla \cdot \Omega) \, \partial_t \Omega 
\, + \, \lambda'_7 \, \rho \, {\mathcal O}_\bot (\Omega \cdot \nabla) \, \partial_t \Omega 
\label{express_T}
\end{eqnarray}
with
\begin{eqnarray*}
& & \hspace{-1cm}
\lambda'_1 = \lambda^1_{11} \, , \quad 
\lambda'_2 = -\lambda^1_{11} + \lambda^2_{11} - \lambda_{21}  \, , \quad 
\lambda'_3 =  \lambda^1_{12}  \, ,  \quad 
\lambda'_4 = \frac{1}{2} \lambda^1_{12} - \lambda_{23},
\\
& & \hspace{-1cm}
\lambda'_5 = - \frac{1}{2} \lambda^1_{12}\, , \quad
\lambda'_6 = \frac{1}{2} \lambda^1_{12} + \lambda^2_{12} - \lambda_{22}, \quad 
\lambda'_7 = \lambda^1_{12} . 
\end{eqnarray*}
Now, we use that up to order $\varepsilon$ terms, we have: 
\begin{eqnarray*} 
& & \hspace{-1cm}
\partial_t \rho =-  \nabla \cdot (c_1 \rho \Omega)  , \\
& & \hspace{-1cm}
  \rho \,  \partial_t \Omega  = -  c_2 \rho \, (\Omega \cdot \nabla) \Omega  -  c_3  \,  (\mbox{Id} - \Omega \otimes \Omega) \nabla \rho ,
\end{eqnarray*}
and replace the time derivatives appearing in (\ref{express_T}) by space derivatives. Using that
\begin{eqnarray}
& & \hspace{-1cm}
{\mathcal O}_\bot (\nabla \Omega) \nabla\rho  = (\nabla \Omega)_{\bot,\bot} \nabla_\bot \rho  = \frac{1}{2} (\sigma(\Omega) + \Gamma(\Omega)) \nabla_\bot \rho + \frac{1}{2} (\nabla \cdot \Omega) \nabla_\bot \rho \, , \label{Obot_nabla_Omega_nabla_rho} \\ 
& & \hspace{-1cm}
{\mathcal O}_\bot (\Omega \cdot \nabla) \nabla \rho = {\mathcal O}_\bot (\Omega \cdot \nabla) \nabla_\bot \rho + (\Omega \cdot \nabla \rho) (\Omega \cdot \nabla) \Omega \, , \label{Obot_Omega_nabla_nabla_rho}\\
& & \hspace{-1cm}
\rho {\mathcal O}_\bot (\Omega \cdot \nabla) \frac{\nabla_\bot \rho}{\rho} = 
{\mathcal O}_\bot (\Omega \cdot \nabla) \nabla_\bot \rho - \frac{1}{\rho} (\Omega \cdot \nabla \rho) \nabla_\bot \rho \, . 
\end{eqnarray}
We then get:
\begin{eqnarray}
& & \hspace{-1.4cm}
T = \lambda''_1 \, (\nabla \cdot \Omega) \nabla_\bot \rho 
\, + \, \lambda''_2  \, \rho \nabla_\bot (\nabla \cdot \Omega)
\, + \, \lambda''_3 \, \sigma (\Omega) \nabla_\bot \rho 
\nonumber \\
& & \hspace{-1cm}
\, + \, \lambda''_4  \, \Gamma (\Omega) \nabla_\bot \rho
\, + \, \lambda''_5  \, {\mathcal O}_\bot (\Omega \cdot \nabla) \nabla_\bot \rho
\, + \, \lambda''_6  \, (\Omega \cdot \nabla \rho) (\Omega \cdot \nabla) \Omega
\nonumber \\
& & \hspace{-1cm}
\, + \, \lambda''_7 \,  \frac{1}{\rho} (\Omega \cdot \nabla \rho) \nabla_\bot \rho
\, + \, \lambda''_8  \, \rho (\nabla \cdot \Omega) (\Omega \cdot \nabla) \Omega
\, + \, \lambda''_9  \, \rho  \sigma (\Omega) (\Omega \cdot \nabla) \Omega
\nonumber \\
& & \hspace{-1cm}
\, + \, \lambda''_{10}  \, \rho  \Gamma (\Omega) (\Omega \cdot \nabla) \Omega
\, + \, \lambda''_{11}  \, \rho  {\mathcal O}_\bot (\Omega \cdot \nabla)( (\Omega \cdot \nabla) \Omega) \, ,  
\label{express2_T}
\end{eqnarray}
with: 
\begin{eqnarray*}
& & \hspace{-1cm}
\lambda''_1 = - \lambda'_1 \frac{3c_1}{2} - \lambda'_6 c_3 \, , \quad 
\lambda''_2 = -  \lambda'_1 c_1 \, , \quad 
\lambda''_3 = - \lambda'_1 \frac{c_1}{2} - \lambda'_4 c_3 \, , \quad 
\\
& & \hspace{-1cm}
\lambda''_4 = - \lambda'_1 \frac{c_1}{2} - \lambda'_5 c_3 \, , \quad 
\lambda''_5 = - \lambda'_1 c_1 - \lambda'_7 c_3\, , \quad 
\lambda''_6 = - \lambda'_1 c_1 - \lambda'_2 c_2 - \lambda'_3 c_1 \, , \quad 
\\
& & \hspace{-1cm}
\lambda''_7 = - \lambda'_2 c_3 + \lambda'_7 c_3 \, , \quad 
\lambda''_8 = - \lambda'_3 c_1 - \lambda'_6 c_2 \, , \quad 
\lambda''_9 = - \lambda'_4 c_2 \, , \quad 
\\
& & \hspace{-1cm}
\lambda''_{10} = - \lambda'_5 c_2  \, , \quad 
\lambda''_{11} = - \lambda'_7 c_2  \, . \quad 
\end{eqnarray*}

Now, we turn towards the term involving $\omega \cdot \nabla G$: 
\begin{eqnarray*}
& & \hspace{-1cm}
S= - {\mathcal O}_\bot \int_{{\mathbb S}^2} (\omega \cdot \nabla) G \, h \, \omega \, d \omega.
\end{eqnarray*}
We decompose
\begin{eqnarray*}
S &=& - {\mathcal O}_\bot \nabla \cdot \left( \int_{{\mathbb S}^2}  G \, h \, \omega \otimes \omega \, d \omega \right) + {\mathcal O}_\bot  \left( \int_{{\mathbb S}^2}  G \, h' \, \omega \otimes \omega  \otimes \omega\, d \omega \right) (\nabla \Omega) \\
&=& S_1 + S_2, 
\end{eqnarray*}
where the second term has the following meaning: 
\begin{eqnarray*}
& & \hspace{-1cm}
\left( \left( \int_{{\mathbb S}^2}  G \, h' \, \omega \otimes \omega  \otimes \omega\, d \omega \right) (\nabla \Omega) \right)_i = \left( \int_{{\mathbb S}^2}  G \, h' \, \omega \otimes \omega  \otimes \omega\, d \omega \right)_{ijk} (\nabla \Omega)_{jk}, 
\end{eqnarray*}
and Einstein's summation convention is assumed. We again decompose $S_1$ into parallel and normal components: 
\begin{eqnarray*}
S_1  &=& - {\mathcal O}_\bot \nabla \cdot \left( \int_{{\mathbb S}^2}  G_e \, h \,  ({\mathcal O}_\bot \omega) \otimes ({\mathcal O}_\bot \omega) \, d \omega \right)  - {\mathcal O}_\bot \nabla \cdot \left( \int_{{\mathbb S}^2}  G_o \, h \, (\omega \cdot \Omega) \, \Omega \otimes ({\mathcal O}_\bot \omega) \, d \omega \right) \\
& & - {\mathcal O}_\bot \nabla \cdot \left( \int_{{\mathbb S}^2}  G_o \, h \, (\omega \cdot \Omega) \, ({\mathcal O}_\bot \omega) \otimes \Omega \, d \omega \right) - {\mathcal O}_\bot \nabla \cdot \left( \int_{{\mathbb S}^2}  G_e \, h \, (\omega \cdot \Omega)^2 \, \Omega \otimes \Omega \, d \omega \right) \\
& = & S_1^1 + \ldots + S_1^4, 
\end{eqnarray*}
where again, we have used (\ref{odd_tens_pow}) to restrict to the even ($G_e$) or odd ($G_o$) components of $G$ with respect to $({\mathcal O}_\bot \omega)$. Using similar computations as for (\ref{BigO_bot_nabla_Omega}) and (\ref{T21}), we find
\begin{eqnarray*}
& & \hspace{-0.3cm}
- \int_{{\mathbb S}^2}  G_e \, h \,  ({\mathcal O}_\bot \omega) \otimes ({\mathcal O}_\bot \omega) \, d \omega = 
\eta_{11}^1 \, (\Omega \cdot \nabla \rho) \, {\mathcal O}_\bot 
+ \eta_{12}^1 \,  \rho \,  (\nabla \cdot \Omega) \,  {\mathcal O}_\bot 
+  \eta_{13}^1 \,  \rho \,  \sigma(\Omega) , 
\end{eqnarray*}
with 
\begin{eqnarray*}
& & \hspace{-1cm}
\eta_{11}^1= \langle \frac{1}{2} \sin^2 \theta \, a_\parallel \, h \rangle_{M_\Omega}, \quad 
\eta_{12}^1= \langle \frac{1}{4} \sin^4 \theta \, b_1 \, h  + \frac{1}{2} \sin^2 \theta \, b_2 \, h \rangle_{M_\Omega}, \\
& & \hspace{-1cm}
\eta_{13}^1 = \langle \frac{1}{8} \sin^4 \theta \, b_1 \, h  \rangle_{M_\Omega}. 
\end{eqnarray*}
We now note that
\begin{eqnarray}
& & \hspace{-1cm}
{\mathcal O}_\bot \nabla \cdot ((\Omega \cdot \nabla \rho) {\mathcal O}_\bot) =  {\mathcal O}_\bot ( \nabla \Omega ) \nabla \rho + {\mathcal O}_\bot (\Omega\cdot \nabla) \nabla \rho - (\Omega \cdot \nabla \rho ) (\Omega \cdot \nabla) \Omega ,\nonumber
\\
& & \hspace{-1cm}
{\mathcal O}_\bot \nabla \cdot ( \rho (\nabla \cdot \Omega) {\mathcal O}_\bot) = \nabla_\bot \rho (\nabla \cdot \Omega) + \rho \nabla_\bot (\nabla \cdot \Omega) - \rho (\nabla \cdot \Omega) (\Omega \cdot \nabla) \Omega , \nonumber 
\\
& & \hspace{-1cm}
{\mathcal O}_\bot \nabla \cdot ( \rho \sigma( \Omega) ) = \sigma( \Omega) \nabla_\bot \rho + \rho {\mathcal O}_\bot ( \nabla \cdot \sigma( \Omega) ) .  \label{Obot_nabla_Omega_nabla_rho_Obot}
\end{eqnarray}
But then, with (\ref{Obot_nabla_Omega_nabla_rho}) and (\ref{Obot_Omega_nabla_nabla_rho}), (\ref{Obot_nabla_Omega_nabla_rho_Obot}) gives
\begin{eqnarray*}
& & \hspace{-0.4cm}
{\mathcal O}_\bot \nabla \cdot ((\Omega \cdot \nabla \rho) {\mathcal O}_\bot) = 
\frac{1}{2} (\sigma(\Omega) + \Gamma(\Omega)) \nabla_\bot \rho + \frac{1}{2} (\nabla \cdot \Omega) \nabla_\bot \rho + {\mathcal O}_\bot (\Omega \cdot \nabla) \nabla_\bot \rho . 
\end{eqnarray*}
Collecting the above identities, we find
\begin{eqnarray*}
& & 
S_1^1 = (\frac{1}{2} \eta_{11}^1 
+ \eta_{13}^1) \, \sigma(\Omega) \nabla_\bot \rho 
+ \frac{1}{2} \eta_{11}^1  \, \Gamma(\Omega) \nabla_\bot \rho 
+ (\frac{1}{2} \eta_{11}^1 + \eta_{12}^1)  \, (\nabla \cdot \Omega) \nabla_\bot \rho 
\\
& & \hspace{0cm}
+ \eta_{11}^1  \, {\mathcal O}_\bot (\Omega \cdot \nabla) \nabla_\bot \rho 
+ \eta_{12}^1  \, \rho \nabla_\bot (\nabla \cdot \Omega) 
- \eta_{12}^1  \, \rho (\nabla \cdot \Omega) (\Omega \cdot \nabla) \Omega 
+ \eta_{13}^1  \, \rho {\mathcal O}_\bot ( \nabla \cdot \sigma( \Omega) ) .
\end{eqnarray*}
Now, turning to $S_1^2$ we have: 
\begin{eqnarray*}
& & \hspace{-1cm}
- \int_{{\mathbb S}^2}  G_o \, h \, (\omega \cdot \Omega) \, \Omega \otimes ({\mathcal O}_\bot \omega) \, d \omega = 
\eta_{11}^2 \, \Omega \otimes \nabla_\bot \rho 
+ \eta_{12}^2 \,  \rho \, \Omega \otimes (\Omega \cdot \nabla) \Omega , 
\end{eqnarray*}
with 
\begin{eqnarray}
& & \hspace{-1cm}
\eta_{11}^2= \langle \frac{1}{2} \sin^2 \theta \, \cos \theta \, a_\bot \, h \rangle_{M_\Omega}, \quad 
\eta_{12}^2= \langle \frac{1}{2} \sin^2 \theta \, \cos \theta \, b_\parallel \, h \rangle_{M_\Omega}. 
\label{mu112_mu122}
\end{eqnarray}
Noting that 
\begin{eqnarray*}
& & \hspace{-1cm}
{\mathcal O}_\bot \nabla \cdot (\Omega \otimes \nabla_\bot \rho)  =  (\nabla \cdot \Omega) \nabla_\bot \rho 
+  {\mathcal O}_\bot (\Omega \cdot \nabla) \nabla_\bot \rho ,
\\
& & \hspace{-1cm}
{\mathcal O}_\bot \nabla \cdot (\Omega \otimes (\Omega \cdot \nabla) \Omega) = 
(\nabla \cdot \Omega) (\Omega \cdot \nabla) \Omega 
+ {\mathcal O}_\bot (\Omega \cdot \nabla)( (\Omega \cdot \nabla) \Omega),
\end{eqnarray*}
we get
\begin{eqnarray*}
& & \hspace{-1cm}
S_1^2 = \eta_{11}^2  \, (\nabla \cdot \Omega) \nabla_\bot \rho 
+ \eta_{11}^2  \, {\mathcal O}_\bot (\Omega \cdot \nabla) \nabla_\bot \rho
\\
& & \hspace{0cm}
+ \eta_{12}^2  \, \rho (\nabla \cdot \Omega) (\Omega \cdot \nabla) \Omega 
+ \eta_{12}^2  \, \rho {\mathcal O}_\bot (\Omega \cdot \nabla)( (\Omega \cdot \nabla) \Omega) .
\end{eqnarray*}
For $S_1^3$, we have 
\begin{eqnarray*}
& & \hspace{-1cm}
- \int_{{\mathbb S}^2}  G_o \, h \, (\omega \cdot \Omega) \, ({\mathcal O}_\bot \omega) \otimes \Omega \, d \omega = 
\eta_{11}^2 \, \nabla_\bot \rho \otimes \Omega 
+ \eta_{12}^2 \,  \rho \, ((\Omega \cdot \nabla) \Omega) \otimes \Omega , 
\end{eqnarray*}
with $\eta_{11}^2$ and $\eta_{12}^2$ given by (\ref{mu112_mu122}). With 
\begin{eqnarray*}
& & \hspace{-1cm}
{\mathcal O}_\bot \nabla \cdot (\nabla_\bot \rho \otimes \Omega)  =  ( \nabla_\bot \rho \cdot \nabla ) \Omega = (\nabla \Omega)^T \nabla_\bot \rho = (\nabla \Omega)_{\bot,\bot}^T \nabla_\bot \rho \\
& & \hspace{3cm}
 = \frac{1}{2} \sigma(\Omega) \nabla_\bot \rho - \frac{1}{2} \Gamma (\Omega) \nabla_\bot \rho + \frac{1}{2} (\nabla \cdot \Omega) \nabla_\bot \rho , \\
& & \hspace{-1cm}
{\mathcal O}_\bot \nabla \cdot (((\Omega \cdot \nabla) \Omega) \otimes \Omega) = 
{\mathcal O}_\bot (\Omega \cdot \nabla)( (\Omega \cdot \nabla) \Omega),
\end{eqnarray*}
we find 
\begin{eqnarray*}
& & \hspace{-1cm}
S_1^3 = \eta_{11}^2  \, ( \frac{1}{2} \sigma(\Omega) \nabla_\bot \rho - \frac{1}{2} \Gamma (\Omega) \nabla_\bot \rho + \frac{1}{2} (\nabla \cdot \Omega) \nabla_\bot \rho ) 
+ \\
& & \hspace{4cm}
+ \eta_{12}^2  \, \rho {\mathcal O}_\bot (\Omega \cdot \nabla)( (\Omega \cdot \nabla) \Omega).
\end{eqnarray*}
Then, for $S_1^4$, we write: 
\begin{eqnarray*}
& & \hspace{-1cm}
- \int_{{\mathbb S}^2}  G_o \, h \, (\omega \cdot \Omega)^2 \, \, d \omega = 
\eta_{11}^4 \, (\Omega \cdot \nabla \rho)
+ \eta_{12}^4 \,  \rho \, (\nabla \cdot \Omega)  , 
\end{eqnarray*}
with $\eta_{11}^4$ and $\eta_{12}^4$ given by 
\begin{eqnarray*}
& & \hspace{-1cm}
\eta_{11}^4= \langle \cos^2 \theta \, a_\parallel \, h \rangle_{M_\Omega}, \quad 
\eta_{12}^4= \langle ( \frac{1}{2} \sin^2 \theta \, b_1 + b_2) \cos^2 \theta \, h \rangle_{M_\Omega}. 
\end{eqnarray*}
With 
\begin{eqnarray*}
& & \hspace{-1cm}
{\mathcal O}_\bot \nabla \cdot ((\Omega \cdot \nabla \rho) \Omega \otimes \Omega)  =  (\Omega \cdot \nabla \rho) (\Omega \cdot \nabla) \Omega, 
\\
& & \hspace{-1cm}
{\mathcal O}_\bot \nabla \cdot (\rho \, (\nabla \cdot \Omega) \Omega \otimes \Omega) = 
\rho (\nabla \cdot \Omega) (\Omega \cdot \nabla) \Omega,
\end{eqnarray*}
we find 
\begin{eqnarray*}
& & \hspace{-1cm}
S_1^4 = \eta_{11}^4  \, (\Omega \cdot \nabla \rho) (\Omega \cdot \nabla) \Omega
+ \eta_{12}^4  \, \rho (\nabla \cdot \Omega) (\Omega \cdot \nabla) \Omega.
\end{eqnarray*}
Now, we turn to $S_2$. Using that $(\nabla \Omega) \Omega = 0$, the decomposition of $\omega$ into ${\mathcal O}_\bot \omega$ and $(\omega \cdot \Omega) \Omega$ reduces to: 
\begin{eqnarray*}
& & \hspace{-1cm}
S_2 = \left( \int_{{\mathbb S}^2}  G_o \, h' \, {\mathcal O}_\bot  \omega \otimes {\mathcal O}_\bot \omega  \otimes {\mathcal O}_\bot \omega\, d \omega \right)_{ijk} (\nabla \Omega)_{jk} +
\\
& & \hspace{1cm}
+ \left( \int_{{\mathbb S}^2}  G_e \, h' \, (\omega \cdot \Omega) \,  {\mathcal O}_\bot  \omega \otimes \Omega  \otimes {\mathcal O}_\bot \omega\, d \omega \right)_{ijk} (\nabla \Omega)_{jk} = S_2^1 + S_2^2, 
\end{eqnarray*}
(where again, Einstein's summation convention has been used). Using (\ref{BigO_bot_nabla_Omega}), we find
\begin{eqnarray*}
& & \hspace{-0.4cm}
-S_2^1 = 
\eta_{21}^1  \, (\sigma (\Omega) \nabla_\bot \rho + 2 (\nabla \cdot \Omega) \nabla_\bot \rho )
+ \eta_{22}^1  \, \rho (\sigma (\Omega) (\Omega \cdot \nabla) \Omega + 2 (\nabla \cdot \Omega) (\Omega \cdot \nabla) \Omega) , \\
& & \hspace{-0.4cm}
-S_2^2 = 
\eta_{21}^2  \, (\Omega \cdot \nabla \rho) (\Omega \cdot \nabla) \Omega 
+ \eta_{22}^2  \, \rho \sigma (\Omega) ((\Omega \cdot \nabla) \Omega) 
+ \eta_{23}^2 \rho (\nabla \cdot \Omega) (\Omega \cdot \nabla) \Omega , 
\end{eqnarray*}
with 
\begin{eqnarray*}
& & \hspace{-1cm}
\eta_{21}^1= \langle \frac{1}{8} \sin^4 \theta \, a_\bot \, h' \rangle_{M_\Omega}, \quad 
\eta_{22}^1= \langle \frac{1}{8} \sin^4 \theta \, b_\parallel h' \rangle_{M_\Omega}, 
\\
& & \hspace{-1cm}
\eta_{21}^2= \langle \frac{1}{2} \sin^2 \theta \, \cos \theta \, a_\parallel \, h' \rangle_{M_\Omega}, \quad 
\eta_{22}^2= \langle \frac{1}{8} \sin^4 \theta \, \cos \theta \, b_1 \, h' \rangle_{M_\Omega}, 
\\
& & \hspace{-1cm}
\eta_{23}^2= \langle (\frac{1}{4} \sin^4 \theta \, b_1 + \frac{1}{2} \sin^2 \theta \, b_2)\, \cos \theta \, h' \rangle_{M_\Omega}. 
\end{eqnarray*}
Collecting all these identities, we find
\begin{eqnarray}
& & \hspace{-0.5cm}
S = \eta'_1 \, (\nabla \cdot \Omega) \nabla_\bot \rho 
\, + \, \eta'_2  \, \rho \nabla_\bot (\nabla \cdot \Omega)
\, + \, \eta'_3 \, \sigma (\Omega) \nabla_\bot \rho 
\nonumber \\
& & \hspace{-0.2cm}
\, + \, \eta'_4  \, \Gamma (\Omega) \nabla_\bot \rho
\, + \, \eta'_5  \, {\mathcal O}_\bot (\Omega \cdot \nabla) \nabla_\bot \rho
\, + \, \eta'_6  \, (\Omega \cdot \nabla \rho) (\Omega \cdot \nabla) \Omega
\nonumber \\
& & \hspace{-0.2cm}
\, + \, \eta'_8  \, \rho (\nabla \cdot \Omega) (\Omega \cdot \nabla) \Omega
\, + \, \eta'_9  \, \rho  \sigma (\Omega) (\Omega \cdot \nabla) \Omega
\, + \, \eta'_{11}  \, \rho  {\mathcal O}_\bot (\Omega \cdot \nabla)( (\Omega \cdot \nabla) \Omega) \, 
\nonumber \\
& & \hspace{-0.2cm}
\, + \, \eta'_{12}  \, \rho  {\mathcal O}_\bot \nabla \cdot \sigma (\Omega),   
\label{express2_S}
\end{eqnarray}
with
\begin{eqnarray*}
& & \hspace{-0.4cm}
\eta'_1 =  \frac{1}{2} \eta_{11}^1 + \eta_{12}^1 + \frac{3}{2} \eta_{11}^2 -2 \eta_{21}^1 \, , \quad 
\eta'_2 =  \eta_{12}^1 \, , \quad 
\eta'_3 = \frac{1}{2} \eta_{11}^1 + \eta_{13}^1 + \frac{1}{2} \eta_{11}^2 - \eta_{21}^1 \, , \quad 
\\
& & \hspace{-0.4cm}
\eta'_4 = \frac{1}{2} \eta_{11}^1 - \frac{1}{2} \eta_{11}^2\, , \quad 
\eta'_5 = \eta_{11}^1 + \eta_{11}^2 \, , \quad 
\eta'_6 = \eta_{11}^4 - \eta_{21}^2 \, , \quad 
\\
& & \hspace{-0.4cm}
\eta'_8 =  - \eta_{12}^1 + \eta_{12}^2 + \eta_{12}^4 -2 \eta_{22}^1 - \eta_{23}^2 \, , \quad 
\eta'_9 = - \eta_{22}^1 - \eta_{22}^2 \, , \quad 
\eta'_{11} = 2 \eta_{12}^2  \, , \quad 
\eta'_{12} = \eta_{13}^1  \, . \quad 
\end{eqnarray*}

We now turn to the last term 
\begin{eqnarray*}
& & \hspace{-1cm}
U = - {\mathcal O}_\bot \int_{{\mathbb S}^2} \nabla_{\omega} \cdot (F_2 \rho M_{\Omega}) \, h \, \omega \, d \omega := {\mathcal O}_\bot \tilde U.
\end{eqnarray*}
The $k$-th component $\tilde U \cdot e_k$ of $\tilde U$ in a Cartesian basis $(e_k)_{k=1,2,3}$ can be transformed by using Stokes theorem on the sphere: 
\begin{eqnarray*}
& & \hspace{-1cm}
\tilde U \cdot e_k = - \int_{{\mathbb S}^2} \nabla_{\omega} \cdot (F_2 \rho M_{\Omega}) \, h \, (\omega \cdot e_k) \, d \omega
= \int_{{\mathbb S}^2} \rho M_{\Omega} \, (F_2  \cdot \nabla_{\omega}) (h (\omega \cdot e_k)) \, d \omega.
\end{eqnarray*}
An easy computation gives
\begin{eqnarray*}
& & \hspace{-1cm}
(F_2 \cdot \nabla_{\omega}) (h (\omega \cdot e_k)) = (\omega \cdot e_k) h' (F_2  \cdot \Omega) + h (F_2\cdot e_k) .  
\end{eqnarray*}
Therefore, 
\begin{eqnarray}
& & \hspace{-1cm}
U = U_1 + U_2, \nonumber \\
& & \hspace{-1cm}
U_1 = \rho  \, {\mathcal O}_\bot \int_{{\mathbb S}^2} M_{\Omega} \, (F_2  \cdot \Omega) \, h' \, \omega \, d \omega , 
\quad U_2 =  \rho  \, {\mathcal O}_\bot  \int_{{\mathbb S}^2} M_{\Omega} \, h  \, F_2 \,d \omega. 
\label{U1_U2}
\end{eqnarray}
From (\ref{Fepsilon_expan}), we can write $F_2 = F_2^1 + F_2^2$. Introducing this decomposition into the expressions (\ref{U1_U2}) of $U_1$ and $U_2$, we get
\begin{eqnarray*}
& & \hspace{-1cm}
U_1 = U_1^1 + U_1^2, \quad  U_2 = U_2^1 + U_2^2, 
\end{eqnarray*}
where for instance, $U_1^2$ is defined by the first expression (\ref{U1_U2}) with $F_2$ substituted by $F_2^2$. In each of the expressions defining $U_p^q$ with $p,q = 1,2$,  we decompose $\omega$ into ${\mathcal O}_\bot \omega + (\omega \cdot \Omega) \Omega$ and, using (\ref{odd_tens_pow}), keep only the even powers of ${\mathcal O}_\bot \omega$. We find: 
\begin{eqnarray*}
& & \hspace{-1cm}
U_p^q = \xi_p^q \, \rho {\mathcal O}_\bot \bar \omega_2,
\end{eqnarray*}
with
\begin{eqnarray*}
& & \hspace{-1cm}
\xi_1^1 = - \langle \frac{1}{2} \sin^2 \theta \, \cos \theta \, \nu \, h' \rangle_{M_\Omega} , \quad 
\xi_1^2 = \langle \frac{1}{2} \sin^4 \theta \, \nu' \, h' \rangle_{M_\Omega} , \quad 
\\
& & \hspace{-1cm}
\xi_2^1 =  \langle (1 - \frac{1}{2} \sin^2 \theta) \, \nu \, h \rangle_{M_\Omega} , \quad 
\xi_2^2 = - \langle \frac{1}{2} \sin^2 \theta \, \cos \theta \, \nu' \, h \rangle_{M_\Omega} , \quad 
\end{eqnarray*}
Therefore, using (\ref{bar_omega_2}) and the fact that $j = c_1 \rho \Omega$ (see Ref. \cite{Degond_M3AS08}), we have:
\begin{eqnarray*}
& & \hspace{-1cm}
U = \xi \, {\mathcal O}_\bot \Delta (\rho \Omega), \quad \xi = {\mathcal K} (\xi_1^1 + \xi_1^2 + \xi_2^1 + \xi_2^2). 
\end{eqnarray*}
Next, we decompose:
\begin{eqnarray*}
& & \hspace{-1cm}
{\mathcal O}_\bot \Delta (\rho \Omega) = {\mathcal O}_\bot (2 (\nabla \Omega)^T \nabla \rho + \rho \nabla \cdot (\nabla \Omega)) . 
\end{eqnarray*}
Now, with (\ref{nabla_botbot_Omega}) and  (\ref{decomp_nabla_Omega_1}), we have:
\begin{eqnarray*}
& & \hspace{-0.4cm}
{\mathcal O}_\bot (\nabla \Omega)^T \nabla \rho = 
\frac{1}{2} \sigma(\Omega) \nabla_\bot \rho - \frac{1}{2} \Gamma(\Omega) \nabla_\bot \rho + \frac{1}{2} (\nabla \cdot \Omega) \nabla_\bot \rho + (\Omega \cdot \nabla \rho) (\Omega \cdot \nabla) \Omega, \\
& & \hspace{-0.4cm}
{\mathcal O}_\bot \nabla \cdot (\nabla \Omega) = 
\frac{1}{2} {\mathcal O}_\bot \nabla \cdot \sigma(\Omega) + \frac{1}{2} {\mathcal O}_\bot \nabla \cdot \Gamma (\Omega) + \frac{1}{2} {\mathcal O}_\bot \nabla \cdot ((\nabla \cdot \Omega) {\mathcal O}_\bot ) \\
& & \hspace{6cm}
+ {\mathcal O}_\bot \nabla \cdot ( \Omega \otimes (\Omega \cdot \nabla) \Omega ) ,
\\
& & \hspace{-0.4cm}
{\mathcal O}_\bot \nabla \cdot ((\nabla \cdot \Omega) {\mathcal O}_\bot ) = \nabla_\bot (\nabla \cdot \Omega) - (\nabla \cdot \Omega) (\Omega \cdot \nabla) \Omega,
\\
& & \hspace{-0.4cm}
{\mathcal O}_\bot \nabla \cdot ( \Omega \otimes (\Omega \cdot \nabla) \Omega ) = (\nabla \cdot \Omega) (\Omega \cdot \nabla) \Omega + {\mathcal O}_\bot (\Omega \cdot \nabla) (\Omega \cdot \nabla) \Omega. 
\end{eqnarray*}
Collecting these results, we finally get: 
\begin{eqnarray}
& & \hspace{-1.4cm}
U = \xi_1 \, (\nabla \cdot \Omega) \nabla_\bot \rho 
\, + \, \xi_2  \, \rho \nabla_\bot (\nabla \cdot \Omega)
\, + \, \xi_3 \, \sigma (\Omega) \nabla_\bot \rho 
\nonumber \\
& & \hspace{-1cm}
\, + \, \xi_4  \, \Gamma (\Omega) \nabla_\bot \rho
\, + \, \xi_6  \, (\Omega \cdot \nabla \rho) (\Omega \cdot \nabla) \Omega
\, + \, \xi_8  \, \rho (\nabla \cdot \Omega) (\Omega \cdot \nabla) \Omega
\nonumber \\
& & \hspace{-1cm}
\, + \, \xi_{11}  \, \rho  {\mathcal O}_\bot (\Omega \cdot \nabla)( (\Omega \cdot \nabla) \Omega) \, ,  
\, + \, \xi_{12}  \, \rho  {\mathcal O}_\bot \nabla \cdot \sigma (\Omega)   
\, + \, \xi_{13}  \, \rho  {\mathcal O}_\bot \nabla \cdot \Gamma (\Omega),   
\label{express2_U}
\end{eqnarray}
with
\begin{eqnarray*}
& & \hspace{-1cm}
\xi_1 = \xi \, , \quad 
\xi_2 = \frac{1}{2} \xi \, , \quad 
\xi_3 = \xi \, , \quad 
\xi_4 = - \xi  \, , \quad 
\xi_6 = 2 \xi \, , \quad 
\xi_8 = \frac{1}{2} \xi \, , \quad 
\\
& & \hspace{-1cm}
\xi_{11} = \xi  \, , \quad 
\xi_{12} = \frac{1}{2} \xi  \, , \quad 
\xi_{13} = \frac{1}{2} \xi  \, . \quad 
\end{eqnarray*}

We can now collect (\ref{express2_T}), (\ref{express2_S}), (\ref{express2_U}) and insert them into (\ref{express_rhs_FP_Omega}) and find  
\begin{eqnarray*} 
& & \hspace{-1cm}  
R_2 = \zeta_1 \, (\nabla \cdot \Omega) \nabla_\bot \rho 
\, + \, \zeta_2  \, \rho \nabla_\bot (\nabla \cdot \Omega)
\, + \, \zeta_3 \, \sigma (\Omega) \nabla_\bot \rho 
\nonumber \\
& & \hspace{-0.5cm}
\, + \, \zeta_4  \, \Gamma (\Omega) \nabla_\bot \rho
\, + \, \zeta_5  \, {\mathcal O}_\bot (\Omega \cdot \nabla) \nabla_\bot \rho
\, + \, \zeta_6  \, (\Omega \cdot \nabla \rho) (\Omega \cdot \nabla) \Omega
\nonumber \\
& & \hspace{-0.5cm}
\, + \, \zeta_7 \,  \frac{1}{\rho} (\Omega \cdot \nabla \rho) \nabla_\bot \rho
\, + \, \zeta_8  \, \rho (\nabla \cdot \Omega) (\Omega \cdot \nabla) \Omega
\, + \, \zeta_9  \, \rho  \sigma (\Omega) (\Omega \cdot \nabla) \Omega
\nonumber \\
& & \hspace{-0.5cm}
\, + \, \zeta_{10}  \, \rho  \Gamma (\Omega) (\Omega \cdot \nabla) \Omega
\, + \, \zeta_{11}  \, \rho  {\mathcal O}_\bot (\Omega \cdot \nabla)( (\Omega \cdot \nabla) \Omega)   
\, + \, \zeta_{12}  \, \rho  {\mathcal O}_\bot \nabla \cdot \sigma (\Omega)   
\nonumber \\
& & \hspace{-0.5cm}
\, + \, \zeta_{13}  \, \rho  {\mathcal O}_\bot \nabla \cdot \Gamma (\Omega) ,  
\end{eqnarray*}
with 
\begin{eqnarray*}
& & \hspace{-1cm}
\zeta_j = \frac{2d}{\langle \sin^2 \theta \nu h \rangle_{M_{\Omega}}} (\lambda''_j + \eta'_j + \xi_j) , \quad j=1, \ldots 13, 
\end{eqnarray*}
and where we have defined the missing coefficients $\lambda''_j$ for $j=12, 13$, $\eta'_j$ for $j = 7, 10, 13$ and $\xi_j$ for $j= 5, 7, 9, 10$ as zero. 

Now, the proof is complete. The expressions of the coefficients ${\mathcal Q}_i$ and ${\mathcal D}_i$ are as follows: 
\begin{eqnarray} 
& & \hspace{-1cm}  
{\mathcal Q}_1 = \frac{1}{\rho} \zeta_7, \, 
{\mathcal Q}_2 = \zeta_1, \, 
{\mathcal Q}_3 = \zeta_3, \, 
{\mathcal Q}_4 = \zeta_4, \, 
\label{expr_mathcal_Q_1}\nonumber \\
& & \hspace{-1cm}
{\mathcal Q}_5 = \zeta_6, \, 
{\mathcal Q}_6 = \rho \zeta_8, \, 
{\mathcal Q}_7 = \rho \zeta_9, \, 
{\mathcal Q}_8 = \rho \zeta_{10}, \, 
\label{expr_mathcal_Q_2} \\
& & \hspace{-1cm}
{\mathcal D}_1 = \zeta_5, \, 
{\mathcal D}_2 = \rho \zeta_{11}, \, 
{\mathcal D}_3 = \rho \zeta_2, \, 
{\mathcal D}_4 = \rho \zeta_{12}, \, 
{\mathcal D}_5 = \rho \zeta_{13}.
\label{expr_mathcal_D} 
\end{eqnarray}
\endproof


\setcounter{equation}{0}
\section{Conclusion}

In this paper, we have provided the $O(\varepsilon)$ corrections to the hydrodynamic model derived in Refs.  \cite{Degond_CRAS07,Degond_M3AS08} from the kinetic description of the Vicsek alignment dynamics\cite{Vicsek_PRL95}. The $O(\varepsilon)$ corrected model involves diffusion terms in both the mass and velocity equations as well as terms which are quadratic functions of the first order derivatives of the density and velocity. To express these terms, it is necessary to decompose the density $\rho$ and velocity $\Omega$ and their gradients in the directions parallel and normal to $\Omega$, thereby expressing that the fluid is non-isotropic about $\Omega$. Future works are concerned with the mathematical theory of this system at least in a simplified form, with the derivation of asymptotic formula for the coefficients in the limits of small and large noise and with numerical simulations and comparisons with the original particle dynamics. In particular, a question to be examined is whether including the $O(\varepsilon)$ corrections in the simulation allows to bypass the ambiguities of the non-conservative hydrodynamic model (see Ref. \cite{Motsch_preprint}) and to yield a better approximation of the solutions of the original particle model


\section*{Acknowledgment}
This work was supported by the Marie Curie Actions of the European 
Commission in the frame of the DEASE project (MEST-CT-2005-021122), by the french 'Agence Nationale pour la Recherche (ANR)' in the frame of the contract 'Panurge' (ANR-07-BLAN-0208-03) and by the General Research Fund of Hong-Kong, CityU 103109.

\end{document}